\documentclass[usenatbib,letter]{mn2e} 
\usepackage{amsmath} 
\usepackage{amssymb} 
\usepackage{graphics}
\usepackage{epsfig}  
\def\be{\begin{equation}}
\def\ee{\end{equation}}
\def\ba{\begin{eqnarray}}
\def\ea{\end{eqnarray}}

\usepackage{color}
\definecolor{red}{rgb}{1,0.0,0.0}

\usepackage[normalem]{ulem}
\definecolor{darkgreen}{rgb}{0.0,0.5,0.0}

\newcommand{\LCDM}{$\Lambda$CDM~}
\newcommand{\beq}{\begin{eqnarray}}  
\newcommand{\eeq}{\end{eqnarray}}  
  
\newcommand{\apj}{ApJ}  
\newcommand{\apjs}{ApJS}  
\newcommand{\apjl}{ApJL}  
\newcommand{\aj}{AJ}  
\newcommand{\mnras}{MNRAS}  
\newcommand{\aap}{A\&A}

\newcommand{\nat}{Nature}

\newcommand{\pasj}{PASJ}    
\newcommand{\avg}[1]{\langle{#1}\rangle}  
\newcommand{\hMpc}{{\ifmmode{h^{-1}{\rm Mpc}}\else{$h^{-1}$Mpc }\fi}}  
\newcommand{\hGpc}{{\ifmmode{h^{-1}{\rm Gpc}}\else{$h^{-1}$Gpc }\fi}}  
\newcommand{\hmpc}{{\ifmmode{h^{-1}{\rm Mpc}}\else{$h^{-1}$Mpc }\fi}}  
\newcommand{\hkpc}{{\ifmmode{h^{-1}{\rm kpc}}\else{$h^{-1}$kpc }\fi}}  
\newcommand{\hMsun}{{\ifmmode{h^{-1}{\rm {M_{\odot}}}}\else{$h^{-1}{\rm{M_{\odot}}}$}\fi}}  
\newcommand{\hmsun}{{\ifmmode{h^{-1}{\rm {M_{\odot}}}}\else{$h^{-1}{\rm{M_{\odot}}}$}\fi}}  
\newcommand{\Msun}{{\ifmmode{{\rm {M_{\odot}}}}\else{${\rm{M_{\odot}}}$}\fi}}  
\newcommand{\msun}{{\ifmmode{{\rm {M_{\odot}}}}\else{${\rm{M_{\odot}}}$}\fi}}  
\newcommand{\lya}{{Lyman-$\alpha$ }}
\newcommand{\clara}{{\texttt{CLARA}}~}
\newcommand{\rand}{{\ifmmode{{\mathcal{R}}}\else{${\mathcal{R}}$ }\fi}}  
\begin{document}

\title[Lyman-$\alpha$ escape fraction at high-z with \texttt{CLARA}]{{\texttt{CLARA}}'s view on
  the escape fraction of Lyman-$\alpha$ photons in high redshift galaxies}
\author[Forero-Romero et al.]{
\parbox[t]{\textwidth}{\raggedright 
  Jaime E. Forero-Romero$^1$ \thanks{Email: jforero@aip.de},
  Gustavo Yepes$^2$, 
  Stefan Gottl\"ober$^1$,\\
  Steffen R. Knollmann$^2$, 
  Antonio J. Cuesta$^3$, 
  Francisco Prada$^3$}\\
\vspace*{6pt}\\
$^1$Leibnitz-Institut f\"ur Astrophysik Potsdam (AIP), An der Sternwarte 16, 14482 Potsdam, Germany\\ 
$^2$Grupo de Astrof\'{\i}sica, Universidad Aut\'onoma de Madrid,   Madrid
E-28049, Spain\\
$^3$Instituto de Astrof\'{\i}sica de Andaluc\'{\i}a (CSIC), Camino Bajo de Hu\'etor 50, E-18008, Granada, Spain
}
\maketitle

\begin{abstract}

  Using \clara (Code for Lyman Alpha Radiation Analysis) we constrain
the escape fraction of Lyman-$\alpha$ radiation in galaxies in the
redshift range $5\lesssim z \lesssim 7$, based on the {\em MareNostrum
High-z Universe}, a SPH cosmological simulation with more than 2 billion
particles. We approximate Lyman-$\alpha$ Emitters (LAEs) as dusty
gaseous slabs with \lya radiation sources homogeneously mixed in the
gas.  Escape fractions for such a configuration and for different gas
and dust contents are calculated using our newly developed radiative
transfer code {\texttt{CLARA}}. The results are applied to the {\em
MareNostrum   High-z Universe} numerical galaxies. The model shows a
weak redshift evolution and good agreement with estimations of the
escape fraction as a function of reddening from observations at
$z\sim2.2$ and $z\sim 3$.  We extend the slab model by including
additional dust in a clumpy component in order to reproduce the UV
continuum luminosity function and UV colours at redshifts $z\gtrsim5$.
The LAE Luminosity Function (LF)  based on the extended clumpy model
reproduces broadly the bright end of the LF derived from observations at
$z\sim5$ and $z\sim6$. At $z\sim 7$ our model over-predicts the LF by
roughly a factor of four, presumably because the effects of the neutral
intergalactic medium are not taken into account.  The remaining tension
between the observed and simulated faint end of the LF, both in the
UV-continuum and \lya at redshifts $z\sim 5$ and $z\sim 6$ points
towards an overabundance of simulated LAEs hosted in haloes of masses
$1.0\times10^{10}\hMsun   \leq   M_h \leq 4.0\times 10^{10}\hMsun$.
Given the difficulties in explaining the observed overabundance by dust
absorption, a probable origin of the mismatch are the high star
formation rates in the simulated haloes around the quoted mass range. A
more efficient supernova feedback should be able to regulate the star
formation process in the shallow potential wells of these haloes.

\end{abstract}


\section{Introduction}

The observational study of the early stages of galaxy formation is
starting a golden age.  Among of the best target populations are
galaxies with strong emission in the Lyman-$\alpha$ line, known as Lyman
Alpha Emitters (LAEs) \citep{1967ApJ...147..868P}. Large observational
samples of these galaxies at redshifts $3<z<7$ are already available.
This has allowed the estimation  of luminosity functions (LF) and
angular correlation functions
\citep{1996Natur.382..281H,2002ApJ...568L..75H,2005pgqa.conf..363H,2004AJ....127..563H,1998ApJ...502L..99H,2003AJ....125.1006R,2004ApJ...617L...5M,2006ApJ...648....7K,2006PASJ...58..313S,2009ApJ...706.1136O,2008ApJS..176..301O,2007ApJ...663...10S,2007A&A...474..385N,2008ApJ...677...12O,2009ApJ...696..546S,2011A&A...525A.143C}.
Large samples of high-z LAEs are expected to be gathered in ongoing and
future observations \citep{2008ASPC..399..115H}. 

The importance of LAEs is not only limited to galaxy evolution. The
detailed measurement of their clustering properties, in particular 
the Baryonic Acoustic Oscillations (BAO) feature \citep{2005ApJ...633..560E}, is
expected to be detected at high redshifts, potentially providing useful
constraints on the evolution of dark energy. A key point in this
analysis is understanding the bias of LAEs as tracers of the large scale
structure \citep{2008A&A...487...63W}.

The study of the epoch of reionisation has also greatly benefited from the study
of LAEs, not only because the features of the emission line make its observational
detection unambiguous, but also because the \lya photons are sensitive to the
distribution of neutral hydrogen, and the changes in the line are able to
constrain the ionisation state of the IGM. It is thus of crucial importance to
properly model the propagation of \lya photons through arbitrary gas
distributions which might contain some dust causing the absorption of these
photons \citep{2010ApJ...716..574Z}. 
   
Because of the resonant nature of the line, \lya photons perform a
random walk in space and frequency before escaping the neutral ISM/IGM
and reaching the observer \citep{1973MNRAS.162...43H}. As a result, the
sensitivity of a \lya photon to dust absorption is enhanced. Small
quantities of dust, depending on the amount and dynamical state of the
neutral gas, can significantly diminish the intensity of the \lya line.
The detailed shape of the line profile also depends on the dynamical
state of the gas and its dust content \citep{1991ApJ...370L..85N}. 

The observational estimation of the fraction of \lya photons escaping the ISM
(hereafter \emph{escape fraction}) is challenging. It usually
requires another probe (UV continuum or a non-resonant recombination line such as
H$\alpha$) and some estimation of the continuum dust extinction. Recent constraints at $z\sim
2.2$ are based on blind surveys of \lya and H$\alpha$. As the line ratio
between H$\alpha$ and \lya is constant and given by atomic physics,
the measurements of the  H$\alpha$ line intensity corrected by extinction,
allow for the estimation of the intrinsic \lya emission
\citep{2010Natur.464..562H}.  

Concerning the emission process, there is general agreement that the bulk of
\lya luminosity in high redshift LAEs is triggered by star formation
processes. The fraction of the emission coming from collisional excitation of
the gas is small and there is evidence that AGN activities do not power the \lya line
emission \citep{2004ApJ...608L..21W}.

Due to all the efforts to model LAEs in a cosmological context, it has been
recognised that the predicted abundance of LAEs (when considering the
 intrinsic emission) overestimates by orders of magnitude the
observed abundance as quantified by the luminosity function
\citep{2005MNRAS.357L..11L, 2007ApJ...670..919K, 2010ApJ...716..574Z}. As the number
density of host dark matter haloes is tightly constrained by the allowed range
of cosmological parameters in the \LCDM cosmology, the most
reasonable assumption is that the difference between observed and predicted
abundances is due to the absorption of \lya photons in the ISM. These facts
motivate the crucial importance of a theoretical determination of the escape
fraction of \lya emitting galaxies.

The theoretical models of LAEs populations based on numerical simulations fall
into two families: (i) semi-analytic models and (ii) hydrodynamical models. On
the semi-analytic side, \cite{2005MNRAS.357L..11L} assume a constant escape
fraction for all galaxies, with  values close to $f_{esc}=0.02$. In a more
refined model \cite{2007ApJ...670..919K} allow a variable escape fraction
motivated by a wind feedback model. In the realm of hydrodynamical models the
recent work of \cite{2010ApJ...716..574Z} addresses the effect of the IGM with
full radiative transfer of the \lya line. They find that as a consequence of
pure photon diffusion a LAE can show a low surface brightness, and might be
missed in a survey. This effect introduces an effective escape fraction that
is not the result of dust absorption.

Hydrodynamical simulations of single galaxies \citep{2009ApJ...704.1640L} and
simple gas/dust configurations have been explored as well
\citep{2006A&A...460..397V}. The problem with simulated individual galaxies is
that the small sample available so far does not  allow  for the inference of
useful scalings or valid statistics in a cosmological context. The limitations
of studying simplified configurations is that, even though they allow for a
wide range of models to be simulated, the parameters, such as dust abundance
or star formation rates, are not constrained by any other assumption, making
it impossible to infer possible scalings with galactic properties
already constrained by observations.

\cite{2010MNRAS.402.1449D} have used a low resolution cosmological SPH
simulation to fix the mass contents and star formation rates of the
galaxies. Unfortunately, they only consider the radiative transfer effects of
the \lya as a free parameter in their model, and do not attempt to
bound the escape fraction from physical considerations consistent with the
resonance nature of the \lya\ line.

In this paper, we address the problem of deriving statistics on the
escape fraction of high-z LAEs between redshifts $5\lesssim z\lesssim 7$
due to the effects of the dusty ISM and resonance scattering of the
Lyman-$\alpha$ line within an explicit cosmological context. We base our
physical analysis of the escape fraction for a given dust and gas
abundance on the results of a new state-of-the-art Monte Carlo radiative
transfer code called \clara ({\bf C}ode for {\bf L}yman {\bf A}lpha {\bf
R}adiation {\bf A}nalysis).  The astrophysical application of these
results relies on the  analysis of a SPH galaxy formation  simulation
with 2 billion particles, the {\em MareNostrum High-z   Galaxy Formation
Simulation}. 

We follow the approach of obtaining the escape fraction for a single
family of models (homogeneous slab with different optical depths of gas
and dust), and applying  them  to the galaxies  in  the simulation. The
dust content   has been calculated  in the simulation by matching the
behaviour of the UV continuum (luminosities and colours) with the
observed estimates  at these redshifts \citep{2010MNRAS.403L..31F}.

Our approach is dictated by two main technical constraints: 1) it is
still not feasible to run the radiative transfer code on several
thousands of individual galaxies in the simulation box. 2) the mass and
spatial resolution in our simulation is not high enough for the
radiative transfer calculation to converge on  the escape fraction using
the gas distribution directly from the SPH simulation, according to the
convergence studies by \cite{2009ApJ...704.1640L}. 

Our theoretical approximation for  the gas and source  distributions  is
based on the premise of consistency with the model  we used for dust
extinction, but also with the objective of improving the description of
simulated \lya emitting galaxies by including two features of the
absorption of the \lya line in galaxies that are commonly neglected.
Namely, we consider:  

\begin{itemize}
\item[a)] the
absorption enhancement by the gas content due to the resonant nature of
the line 
\item[ b)] a spatial distribution of the Lyman-alpha regions in
the galaxy where the \lya  photons are not forced to have statistically the
  same probability of being absorbed, as it  is the case for centrally emitted
  \lya\ photons in a sphere, shell or slab configuration.
\end{itemize}

This paper is structured as follows. In Section \ref{sec:galfinder} we
describe the simulation and the galaxy finding technique. In Section
\ref{sec:spectra} we describe our method to calculate the spectral
energy distributions for the galaxies in the sample, as well as our
simplified dust extinction model. We review the  UV continuum properties
of the sample as derived in \cite{2010MNRAS.403L..31F}.  Our model for
LAEs is described in Section \ref{sec:spherical_laes}, together with its
implications on the escape fraction and its implementation into the {\em
MareNostrum High-z Galaxy Formation Simulation}. We discuss the
implications of our model in Section \ref{sec:results}. Finally, we
summarise our conclusions in Section \ref{sec:conclusions}. All the
details regarding the implementation of the \lya Monte Carlo radiative
transfer code  \clara can be found in the Appendix \ref{sec:montecarlo}.

\section{Cosmological Simulation and Galaxy Finding}
\label{sec:galfinder}

The {\em MareNostrum  High-z Universe} simulation{\footnote{\tt
http://astro.ft.uam.es/marenostrum}} follows the non-linear evolution of
structures in baryons (gas and stars) and dark matter, starting  from
$z= 60$  within a cube of $50\hMpc$ comoving on a side.  

The cosmological parameters used are consistent with WMAP1 data
\citep{2003ApJS..148..175S} \emph{i.e.} $\Omega_{\rm m}=0.3$,
$\Omega_{\rm b}=0.045$, $\Omega_\Lambda=0.7$, $\sigma_8=0.9$, Hubble
parameter $h=0.7$, and a spectral index $n=1$. The initial density field
was sampled by $1024^3$ dark matter particles with a mass of $m_{\rm DM}
= 8.2 \times 10^6 \hMsun $ and $1024^3$ SPH gas particles with a mass of
$m_{\rm gas} = 1.4 \times 10^6 \hMsun$. The gravitational smoothing
scale  was set to 2 \hkpc in comoving coordinates.   The simulation has
been performed using the TREEPM+SPH code \texttt{GADGET-2}
\citep{Springel05}.  Further details on the physical setup of the code
can be found in \cite{2010MNRAS.403L..31F}

We identify the objects  in the simulations  using the \texttt{AMIGA}
Halo Finder\footnote{{\tt http://www.popia.ft.uam.es/AMIGA}}
(\texttt{AHF})  which is described in detail in \citet{Knollmann2009}.
\texttt{AHF} takes into account the thermal energy of gas particles
during the calculation of the binding energy. The halo consists only of
bound particles.  All objects with more than  $1000$ particles,  dark
matter, gas and stars combined, are used in our analyses. We assume a
galaxy  is resolved if the object contains $200$ or more stellar
particles,  which  corresponds to objects with $\gtrsim 400$ particles
of gas. This ensures  a proper estimation of the average gas column
densities and star formation rates in the numerical galaxies, in
agreement with recent resolution studies \citep{2010ApJ...711.1198T}.

The favoured cosmological parameters estimated from the analysis of
recent CMB data are different from the ones used in the simulation. We
have included an additional correction in  the galaxy abundance from
the different number density of dark matter haloes in the cosmology used
in the simulation (WMAP1, with $\sigma_8=0.90$
\citealt{2003ApJS..148..175S}) and the values favored in more recent
works   (WMAP5, with $\sigma_8=0.796$ \citealt{2009ApJS..180..306D}).

\section{Spectral Modeling and UV Continuum}
\label{sec:spectra}

In this work, we use the same spectral model and follow the same
extinction model to calculate the Spectral Energy Distribution (SED) for
each galaxy, as described in Section 3  of \cite{2010MNRAS.403L..31F}.
The photometric properties of galaxies are calculated employing the
stellar population synthesis model STARDUST \citep{1999A&A...350..381D},
using the methods described in \cite{2003MNRAS.343...75H}. We adopt a
Salpeter Initial Mass Function (IMF) with lower and upper mass cutoffs
of $0.1$\Msun\ and $120$\Msun. 

The SEDs  are built from the AHF catalogues already described. Each star
particle in the simulation represents a burst of stars of a given
initial mass and metallicity evolved at a given age. We have added all
the individual spectra of each star particle to build UV magnitudes
\citep{2010MNRAS.403L..31F}. This allowed us as well to implement a dust
extinction model in two different stellar populations distinguished by
age as we describe in the next paragraphs. However, we do not use these
SEDs to estimate the production rate of ionizing photons, which is noisy
for galaxies resolved with less than $100000$ particles for physical
reasons detailed in Section \ref{subsection:intrinsic}.

The dust attenuation model parametrises both the extinction in a
homogeneous interstellar medium (ISM) and in the molecular clouds around
young stars, following the physical model of \cite{2000ApJ...539..718C}.
The attenuation from dust in the homogeneous ISM assumes a slab
geometry, while the additional attenuation for young stars is modeled
using spherical symmetry.

We first describe the optical depth for the homogeneous interstellar
medium, denoted by $\tau_{d}^{ISM}(\lambda)$. We take the mean
perpendicular optical depth of a galactic disc at wavelength $\lambda$
to be 

\begin{equation}
\tau_{d}^{ISM}(\lambda)  = \eta
\left(\frac{A_{\lambda}}{A_{V}}\right)_{Z_{\odot}}\left(\frac{Z_g}{Z_{\odot}}\right)^r\left(\frac{\avg{N_{H}}}{2.1
  \times 10^{21} \mathrm{atoms\ cm}^{-2}}\right),
\label{eq:ISM}
\end{equation}
where $A_\lambda/A_V$ is the extinction curve from
\cite{1983A&A...128..212M}, $Z_{g}$ is the gas metallicity, $\avg{N_H}$
is the mean atomic hydrogen column density and $\eta=(1+z)^{-\alpha}$ is
a factor that accounts for the evolution of the dust to gas ratio
at different redshifts,  with $\alpha>0$ from the available constraints
based on simplified  theoretical models \citep{2003PASJ...55..901I} and
observations around $z\sim  3$ \citep{2006ApJ...644..792R}. The
extinction curve depends on the gas metallicity $Z_{g}$ and is based on
an interpolation between the solar neighbourhood and the Large and Small
Magellanic Clouds ($r=1.35$ for $\lambda < 2000 $\AA\ and $r=1.6$ for
$\lambda > 2000$\AA).

The mean Hydrogen column density is calculated as

\begin{equation}
\avg{N_H} = X_H \frac{M_{g}}{m_{p}\pi r_{g}^2}\mathrm{atoms\ cm}^{-2},
\label{eq:column_H}
\end{equation}
where $X_H = 0.75$ is the universal mass fraction of Hydrogen, $M_g$ is the
mass in gas, $r_{g}$ is the radius of the galaxy and $m_p$ is the proton
mass. The radius, stellar and gas masses for each galaxy are taken
from the AHF catalogues, where we have verified that computing $\langle
N_{H}\rangle$ from the galaxy catalogues yields, on average, similar results
than integrating the 3D gas distribution of the galaxies
using the appropriate SPH kernel, provided that the galaxies are
sampled with more than $\sim 200$ gas particles.  

In addition to the foreground, homogeneous ISM extinction, we also
model, in a simple manner,  the attenuation of young stars that are
embedded  in their birth clouds (BC). Stars younger than a given age,
$t_{c}$, are subject to an additional attenuation with mean
perpendicular optical depth

\begin{equation}
\tau_{d}^{BC}(\lambda) = \left(\frac{1}{\mu}-1\right) \tau_{d}^{ISM}(\lambda),
\label{eq:young}
\end{equation}
where $\mu$ is the fraction of the total optical depth for these  young
stars with respect to that is found in the homogeneous ISM.

Without any correction, the simulated LFs have a higher normalisation
than the observational ones, which seems to be a general feature for all
\LCDM  hydrodynamical simulations at high redshift
\citep{2006MNRAS.366..705N}. The excess can be caused by two different
effects, both possibly acting at the same time: the physics included in
the simulation giving rise to excessive star formation rates, or the
intrinsic UV ought to be corrected by dust extinction.  In this section
we review the results from the explanation by a dust correction based on
the physical model described above.

The simple approximation of a dust optical depth proportional to the gas
column density leads to a reddening which scales with the galaxy
luminosity.  Massive and luminous galaxies are more extinguished than
less massive ones.  This is in agreement with similar numerical results
\citep{2006MNRAS.366..705N,2011MNRAS.410.1703F} and observational
constraints at redshift $z\sim 3$ \citep{2001ApJ...562...95S}. Applying
such a correction to the data at redshifts $5<z<7$ cannot explain the
faint end of the LF, because there is still a large overabundance of
simulated galaxies with respect to the observations. However, a constant
reddening $E(B-V)\sim 0.2$ (assuming a Calzetti law) on all galaxies
uniformly dims the LF, thus providing a good match at the faint end.  A
clumpy ISM \citep{2005MNRAS.359..171I} would be a plausible physical
model to explain this effect, which also has  proven to be effective at
high redshift in producing an almost constant reddening as a function of
galaxy luminosity \citep{2010MNRAS.403L..31F}.

The results obtained from matching the observed UV LF between redshifts
$5\leq z\leq 7$ to the LFs derived from the simulation hint that all
stars younger than $25$ Myr must additionally be extinguished with
parameters $\mu=0.01$ for redshifts $z\sim 5,6$ and $\mu=0.03$ for
redshift $z\sim 7$, and $\alpha=1.5$ \citep{2010MNRAS.403L..31F}.  This
means that the young stars have an additional extinction $\mu^{-1} - 1$,
ie.~$30-100$, times larger than the extinction associated to the
homogeneous ISM. Observational constraints at $z=0$ locate $\mu$ around
$1/3$ with a wide range of scatter between $0.1$ and $0.6$
\citep{2004MNRAS.349..769K}.  One possible interpretation  for the
evolution of the $\mu$ parameter is that high redshift galaxies have a
less dust enriched homogeneous ISM  than those at low redshift, making
the relative contribution of  extinction around their young stars
higher.  Within  this interpretation, a factor of  $\sim 30$ increase in
the value of  $\mu$ (with values of $\alpha=1.5$) would require  that
the dust to gas ratio in the homogeneous ISM increases between $z\sim 7$
and $z=0$ by at least a factor of $\sim 30\times (1+z)^\alpha\sim 400$,
which is feasible under conservative theoretical considerations of what
the dust to gas ratio evolution should be. For instance,
\cite{2004ApJ...611...40C} can account for a change of two orders of
magnitude.  However, realistic theoretical estimations of the $\mu$
factor would require simulations of galaxy evolution with spatial
resolution on the order of a few $\sim 100$ pc.
\citep{2010MNRAS.404.2151C}.

Our approach to calculate the dust extinction is thus purely
phenomenological. It does not assume any universal extinction law for
all galaxies and is not based on a dust production model. The extinction
curve for each galaxy is different depending on its metallicity and gas
contents. We can use the scaling between reddening, $E(B-V)$, and
extinction, $A_V$, to benchmark the impact of using a fixed universal
extinction law. The Calzetti extinction curve has
$R_{V}=A_{V}/E(B-V)=4.05$ \citep{2000ApJ...533..682C}, while for a
Supernova (SN) extinction curve it can be $R_{V}=7.8$ or $R_{V}=5.8$
\citep{2005MNRAS.357.1077H}. Results at $z\sim 5$
\citep{2010MNRAS.403L..31F} indicate that for the brightest best
resolved galaxies ($M_{UV}<-21$) $R_{v}\sim 8.0\pm 0.5$ is a fair
approximation for the median values.  This means that for a given amount
of extinction, the expected reddening will be higher for a Calzetti
extinction curve. In other words, one could match the UV magnitudes
using the Calzetti extinction law, as it was done for instance by
\cite{2010MNRAS.403L..84D}, but conflicting results for the UV colours
can be expected in that case.

Our clumpy ISM model, as applied to the UV-continuum, fixes  the optical
depth of dust in the homogeneous and clumpy phases of the IGM. The
hydrogen optical depth in the homogeneous ISM is  fixed by the HI column
densities already calculated, while HI column densities in the clumpy
phase can be bounded by the conditions expected in the molecular clouds
of young star forming regions.  With these constraints we proceed in the
next section to quantify the expected extinction of the \lya\ line based
on a physical model that includes the resonance nature of the line.  In
order to be consistent with the approximation used for the continuum
extinction we also fix the geometry of the gas and dust distribution to
be that of a dusty slab with the radiation sources distributed
homogeneously. In the next sections we will estimate the escape fraction
of \lya\ radiation in such configuration, and illustrate how this
results are consistent with observational constraints of the escape
fraction as a function of reddening.

\section{Slab Approximation for LAEs}
\label{sec:spherical_laes}

We approximate LAEs as homogeneous slabs of gas with dust and \lya
radiation sources homogeneously mixed. The motivation to explore such a
model is twofold. First, we want to be consistent with the
extinction approximation already used for the UV continuum. Second, the
homogeneous distribution keeps an important feature seen in many
simulations, including the best resolved galaxies in the {\em
MareNostrum Simulation} used here, namely that the stars are not
clustered around a single point with respect to the gas.

Concerning the latter point, \cite{2009ApJ...704.1640L} studied the
escape fraction in high resolution simulations of individual galaxies.
The resolution study they performed  indicates that converged values for
the escape fraction require minimal smoothing lengths for the gas on the
order of $160$ $h^{-1}$ pc comoving, which is one order of magnitude
smaller than the  resolution of our {\em MareNostrum High-z Galaxy
Formation Simulation}.

Therefore, it is  still unavoidable to make use of  the  kind of subgrid
models we propose here, in order to derive statistical results on the
escape fraction. However, it is  possible to improve the modeled physics
by considering explicitly the effects of resonant scattering in the line
absorption.

In this Section we will employ {\texttt{CLARA}} to  estimate the escape
fraction in the slab configuration,  assuming  homogeneously mixed
sources. The source distribution constitutes the major  difference
between our work and  similar \lya\ Monte-Carlo radiative transfer
studies \citep{2006MNRAS.367..979H,2006A&A...460..397V}. We will show
that this assumption strongly affects the escape fraction results. We
will then describe how we estimate the relevant physical quantities in
the simulation to obtain an escape fraction for each simulated galaxy.
We show how this model agrees with observational results on the escape
fraction for galaxies at $z\sim 2.2$ and $z\sim 3$.

\subsection{Radiative Transfer Results}

\begin{figure}
\begin{center}
\includegraphics[width=0.45\textwidth]{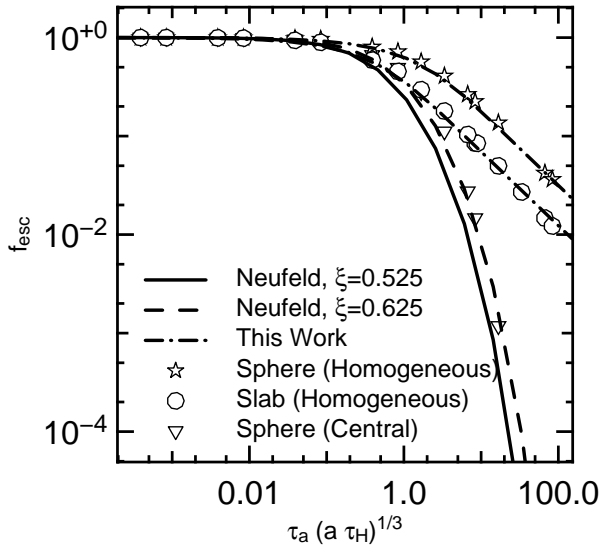}
\end{center}
\caption{Escape fraction for \lya\ photons emitted inside different
  dusty gas configurations as a function of the product $\tau_{a}(a  \tau_{H})^{1/3}$. 
Symbols represent the results obtained with {\texttt{CLARA}}. The
empty triangles show the solution for a dusty
  sphere when the photons are  emitted at the centre of the sphere, whereas stars
  represent the case with \lya sources homogeneously distributed in
  the sphere.  Hexagons represent the results of the infinite dusty
  slab with sources homogeneously distributed.
  The solid line shows the analytical solution  shown in
  Eq. (\ref{eq:f_esc_main}) for  the infinite dusty slab and sources
  located in the slab's centre ($\xi=0.525$).  The
  dashed and   dotted line  represent the same analytical expression with
  parameters  $\xi=0.625$. A better fit for the
  homogeneous  distributed  sources (both for the sphere and the slab)
  is displayed by the dash-dotted line, see Eqs. (\ref{eq:f_esc_homo})
  and (\ref{eq:f_esc_homo_bis}).}  
\label{fig:escape_sphere}
\end{figure}

The problem of an infinite homogeneous dusty gas slab has an analytic
solution for the escape fraction provided that the sources are located
in a thin plane.  The dashed line in Figure \ref{fig:escape_sphere}
corresponds to the theoretical expectation (described in Appendix
\ref{sec:montecarlo}, Eq.(\ref{eq:f_esc})) for the escape fraction out
of a infinite slab as a function of the product $(a\tau_0)^{1/3}\tau_a$,      

\begin{equation}
f_{esc} = \frac{1}{\cosh \left( \xi^{\prime} \sqrt{(a\tau_{0})^{1/3}\tau_a}\right)},
\label{eq:f_esc_main}
\end{equation}
where $\tau_0$ is the
Hydrogen optical depth, $\tau_a$ is the optical depth of absorbing material
(for albedo values of $A$, $\tau_{a}= (1-A)\tau_{d}$, where $\tau_{d}$ is the
dust optical depth), and $a$ is a measure of the temperature in the gas defined as
$a=\Delta\nu_L/(2\Delta\nu_D)$, $\Delta\nu_{D}=(v_{p}/c)\nu_0$ is the Doppler
frequency width, and $v_{p} = (2kT/m_{H})^{1/2}$ is $\sqrt{2}$ times the 
velocity dispersion of the Hydrogen atom, $T$ is the gas temperature, $m_{H}$ is the Hydrogen atom
mass and $\Delta\nu_L$ is the natural line width. The constant
$\xi^{\prime}=\xi\sqrt{3}/\pi^{5/12}$ and  $\xi$ is a free parameter taking a
value of $\xi = 0.525$ in the case of centrally located sources.

For comparison, we show the results in the case of a homogeneous dusty
gas sphere with centrally located sources of \lya radiation. This
configuration is fully parameterised by the optical depth of Hydrogen
and dust as measured from the centre of the sphere to its boundary. We
simulate with \clara a series of spheres with different values for these
optical depths (see Table \ref{table:runs_slab}). The empty triangles in
Figure \ref{fig:escape_sphere} represents the results for that
configuration.   The main point of this numerical experiment is to
confirm that the same function that is used to describe the escape
fraction of the infinite slab seems to be able to describe the escape
fraction out of the sphere in the case of centrally located sources. The
only difference is the value for the constant, in the spherical case
$\xi=0.625$. Nevertheless, the differences between both geometries for
the gas density and the dust abundance, at a given temperature, are
never larger than a factor of 4, for the range of  parameters studied.

We now turn to the case where the sources of \lya radiation are
homogeneously mixed inside the infinite slab. In this configuration we
expect a higher escape fraction given the fact that the photons will be,
on average, emitted closer to the escape surface. This will allow the
photons to be affected by less collisions compared to the photons
emitted at the centre of the slab.  In Figure \ref{fig:escape_sphere},
the escape fraction for the homogeneous distribution of \lya sources is
shown  by the hexagons. For small values of ${\mathcal{T}} \equiv
\tau_{a}(a\tau_{0})^{1/3}$ the escape fraction seems to be well
approximated by Eq.(\ref{eq:f_esc_main}).  For large values of
$\mathcal{T}>1$  the  escape fraction can be up to three orders of
magnitude higher than  in the central emission geometry in the range of
explored values. The scaling of the escape fraction with the variable
$\mathcal{T}$ does not follow Eq.(\ref{eq:f_esc_main}). For large values
of $\mathcal{T}>1$, the escape fraction shows a fall off almost
proportional to $1/\mathcal{T}$. This can be understood in analogy to
the classical case of continuum light emitted inside a dusty slab. As
the optical depth reaches high values, only the photons emitted close to
the surface within a distance $\propto 1/\tau $ have a high escape
probability.

In analogy with the solution of continuum attenuation in a dusty slab, we
find that for \lya radiation, a  solution with the functional form 

\begin{equation}
f_{esc} = \frac{1 - \exp(-P)}{P}
\label{eq:f_esc_homo}
\end{equation}
with 
\begin{equation}
P = \epsilon((a\tau_{0})^{1/3}\tau_a)^{3/4},
\label{eq:f_esc_homo_bis}
\end{equation}
provides a reasonable  description of the Monte-Carlo results as shown
in Figure \ref{fig:escape_sphere} (dot-dashed line) with $\epsilon =
3.5$ in the slab geometry and $\epsilon = 1.0$ in the spherical case.

We conclude that the source distribution, relative to the gas
distribution, has a larger impact on the escape fraction than the
geometrical distribution of the gas itself. The infinite slab
exhibits very similar escape fraction compared to a spherical gas distribution,
even the same scaling with respect to the gas properties holds.  On the
other hand, the homogeneous source distribution presents a radically
different scaling with the optical depth in the gas, allowing for high
escape fractions at large values of $(a\tau_{0})^{1/3}\tau_a > 1$,
both for the slab and spherical geometries. Additionally, the
shape of the outgoing spectra, corrected for the escape fraction are
different for the homogeneous vs.\ central source distribution (see
Appendix \ref{sec:montecarlo} for the case of the dusty sphere).

\subsection{Implementation in the MareNostrum Simulation}
\label{sub:implement}

In the previous subsection, we obtained results on the escape fraction
for the case of a homogeneous dusty gas slab with radiation sources distributed
homogeneously. We provided a fitting formula (Eq. (\ref{eq:f_esc_homo}))
for the escape fraction as a function of the optical depths of gas and
dust and the gas temperature. These results can be used by any
galaxy evolution model that provides predictions for these quantities.

In our case, we use this formula to to derive the values of the escape
fraction for each galaxy in the \emph{MareNostrum High-z Universe}
simulation. In what follows, we explain how we proceed to estimate the
physical quantities we need:  $\tau_{0}$, $\tau_{a}$ and $a$.  

First, we recall from Section \ref{sec:spectra} that the dust model described
above splits the extinction into two contributions:
 
\begin{enumerate}
\item the extinction by the homogeneous ISM on all stars in the galaxy,
\item the extinction by the birth clouds around young stars.
\end{enumerate}

We now want to have estimates for the escape fraction associated with the
homogeneous ISM, $f_{esc}^{ISM}$, and with the birth clouds,
$f_{esc}^{BC}$.  

In order to estimate the ISM escape fraction, $f_{esc}^{ISM}$ we
describe the ISM as a slab of constant density and temperature of
$10^{4}$K, which fixes the value of $a=4.7\times 10^{-4}$.  The hydrogen
column density of this slab is given by  Eq.(\ref{eq:column_H}). The
average optical depth of neutral hydrogen is calculated from this column
density  and the hydrogen cross section  at the center of the Lyman
alpha line at a temperature of $10^4$K.  Based on the results of the
previous subsection, we have thus determined all the parameters we need
to calculate the escape fraction in the slab approximation: the optical
depth of dust  and Hydrogen ($\tau^{ISM}_{d}$, $\tau^{ISM}_{H}$) and the
gas temperature $T$.  

We now have to estimate the escape fraction associated with the birth
clouds, $f_{esc}^{BC}$.  As all the \lya emission comes from these
regions, the full intrinsic \lya luminosity has to be corrected as well
by this escape fraction. The dust optical depth associated with the
clouds, $\tau_d^{BC}$, has already been calculated using
Eq.(\ref{eq:young}). On the other hand, the optical depth of hydrogen in
the birth clouds, $\tau^{BC}_{H}$, still has to be estimated.  Based on
observations of large molecular clouds, the neutral Hydrogen column
density in the densest regions is of the order of $10^{19}$ cm$^{-2}$
\citep{1983ApJ...268..727W}. This is already $\sim 3$ orders of
magnitude lower than the optical depth in our simulated galaxies at high
redshift. Moreover, the warm regions have densities two orders of
magnitudes lower \citep{2001RvMP...73.1031F}. A realistic estimate then
puts the average neutral Hydrogen optical depth $\sim 10^5$ times lower
than the one estimated for the full galaxy.  Under these conditions, we
find within the cloud $\tau^{BC}_{0}\sim \tau^{BC}_{d}$ with $a
\tau^{BC}_{H}<1$ and  $\tau^{BC}_{d}>1$, making the extinction
enhancement by resonant scattering irrelevant.  In this case, we can
take the escape fraction as the continuum extinction at
$\lambda=1260$\AA\ for a spherical geometry.

\subsection{Comparison against observational constraints}

\begin{figure*}
\begin{center}
\includegraphics[width=0.45\textwidth]{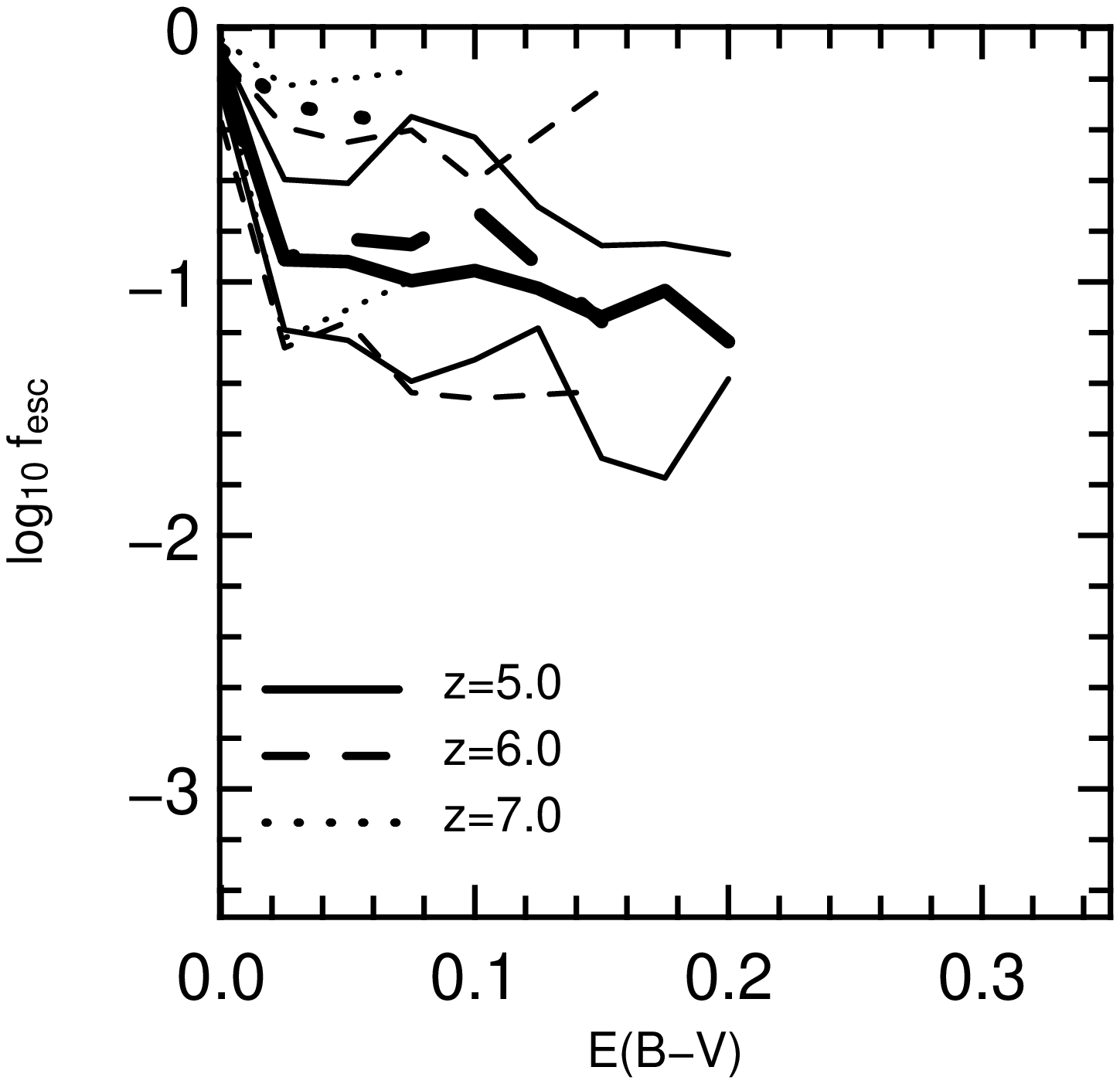}\hspace{1cm}
\includegraphics[width=0.45\textwidth]{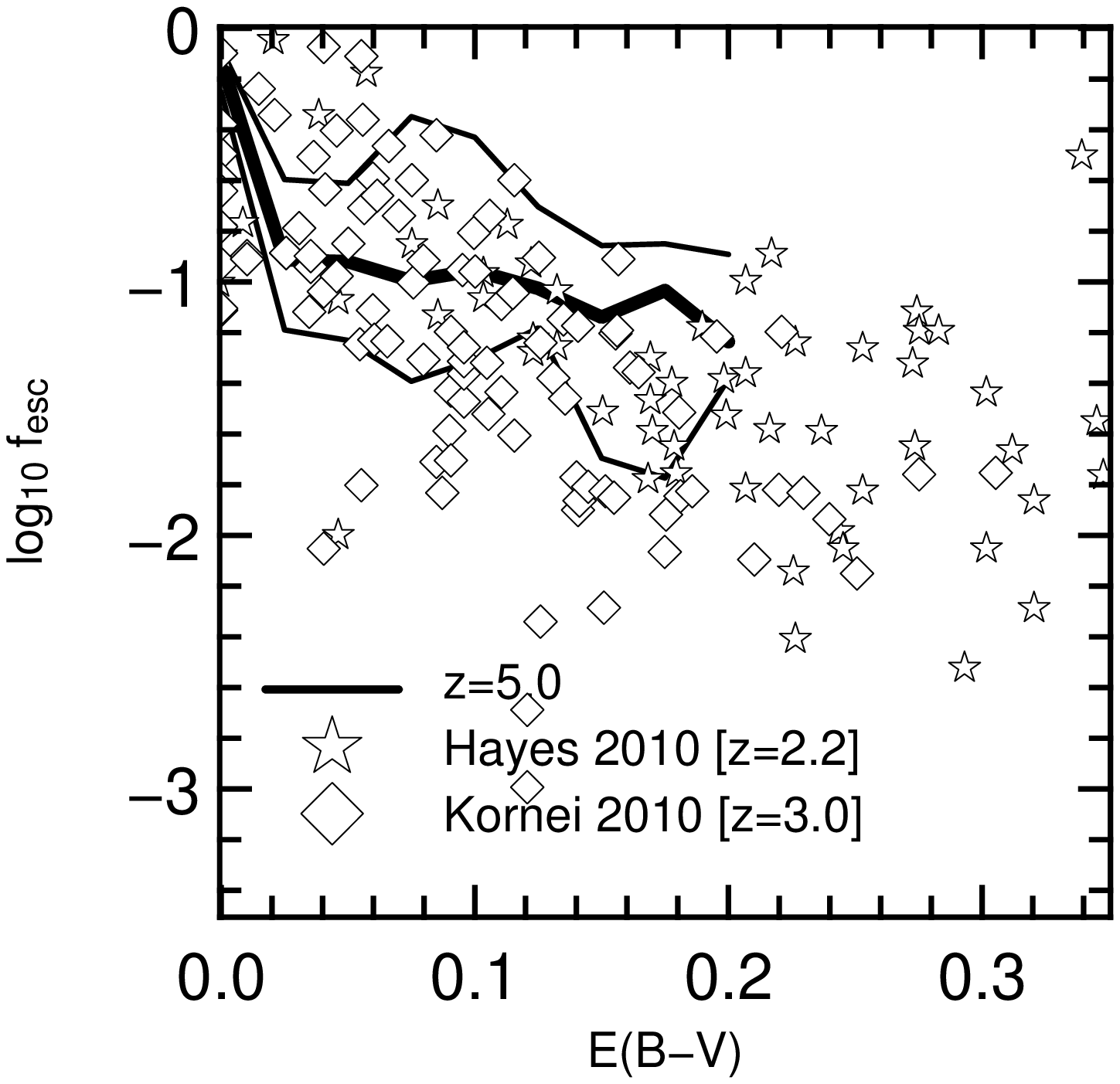}\hspace{1cm}
\end{center}
\caption{Escape fraction as a function of the UV color excess as estimated
  from the homogeneous ISM model. The lines show different distributions obtained
  from the simulation, the thick line marks the median, and the thin lines the
  first and third quartile. Stars represent  observational estimates of the escape fraction at $z\sim 2.2$
  by \citet{2010Natur.464..562H}. Diamonds represent similar estimates
  at $z\sim 3.0$ by \citet{2010ApJ...711..693K}
  In the left panel, the different lines show the evolution between $5<z<7$. We
  observe how to lower redshift values, higher values of the color excess are
  present. The very weak redshift evolution justifies the comparison of the
  results of this model against observations shown in the right panel.  Only the simulated
  galaxies above the resolution threshold were taken into account in
  the calculation. The slab model with homogeneous source distribution
  matches closely the observed scaling of the escape fraction with
  colour excess.  
\label{fig:tau_esc}}  
\end{figure*}

Observational estimates of the escape fraction at redshifts $z\gtrsim 5$
(the lowest current redshift of the \emph{MareNostrum High-z Universe}
simulation) are not available. Nevertheless, at redshifts $z\sim 3$ the
homogeneous ISM model is already consistent with the reddening scaling
with  galaxy luminosity \citep{2001ApJ...562...95S}. It is then
reasonable to think that at all redshifts $z<3$ the homogeneous ISM is
still the appropriate model regarding the extinction. For this reason,
we decide to compare the escape fraction predicted by the homogeneous
ISM with the observational results at $z\sim 2.2$ and $z\sim 3$.

In the left panel of Figure \ref{fig:tau_esc} we present  the redshift
evolution of the escape fraction as a function of reddening at redshifts
$z\sim 5$, $6$, $7$. We find that there is a weak redshift dependence.
The biggest difference at each redshift is the presence of galaxies with
larger values of colour excess.  As the redshift evolution of the
overall scaling is rather weak, we compare our theoretical results at
$z\sim 5$, the last redshift available in the simulation to make this
study, with the observations at  $z\sim 2.2$ and $z\sim 3$, the highest
redshifts so far with observations of this kind.  The physical time
between the observational results and the simulated ones is $\sim 1$Gyr.

In the right panel of Figure \ref{fig:tau_esc} we show the final result of
the comparison between the escape fractions from observations at $z\sim
2.2$ and $z\sim 3$ and our model applied to the galaxies in the
simulation at $z\sim 5$. The empty symbols represent the observational
data, indeed showing a strong scaling: the escape fraction decreases
with increasing colour excess. Our simple model of a dusty gas slab with
homogeneously distributed sources reproduces well these observational
trends determined by \cite{2010Natur.464..562H} and
\cite{2010ApJ...711..693K}.

In order to facilitate the comparison of our results with  those   from
other  works \citep{2010ApJ...711..693K,2010Natur.464..562H}, we also
use the functional form $f_{esc} = C_{Ly\alpha}10^{-0.4\cdot
E_{(B-V)}\cdot k_{Ly\alpha}}$ to fit  the trend of  the median values in
the $f_{esc}$-$E(B-V)$ plane. This  form takes into  account that
$C_{Ly\alpha}\neq 1 $  for   very low extinction values ($E(B-V) \approx
0$),   as it is expected due to the resonance nature of the \lya\ line.
We obtain $C_{Ly\alpha} = 0.21 \pm 0.05$ and $k_{Ly\alpha}=8.6 \pm 1.0$,
which is  somewhat  flatter than the value of $k_{Ly\alpha}=12.0$
obtained by \cite{2000ApJ...533..682C} for \lya\ wavelengths.

\section{Analysis and Results}
\label{sec:results}

In this section we present our results on the escape fraction applying
the slab model to the gas and dust contents calculated from the
\emph{MareNostrum High-z Universe} simulation.  We first derive useful
scalings of the escape fraction with the mass of the DM halo hosting the
galaxy.  We then apply the estimated escape fraction to obtain the
luminosity function of LAEs at high redshift, $z\gtrsim 5$. Given that
at these redshifts the extinction model includes a component associated
to the birth clouds of the stars, we will add this component for the
comparison with the luminosity functions at these redshifts.

\subsection{Scaling with Halo Mass}

\begin{figure*}
\begin{center}
\includegraphics[width=0.45\textwidth]{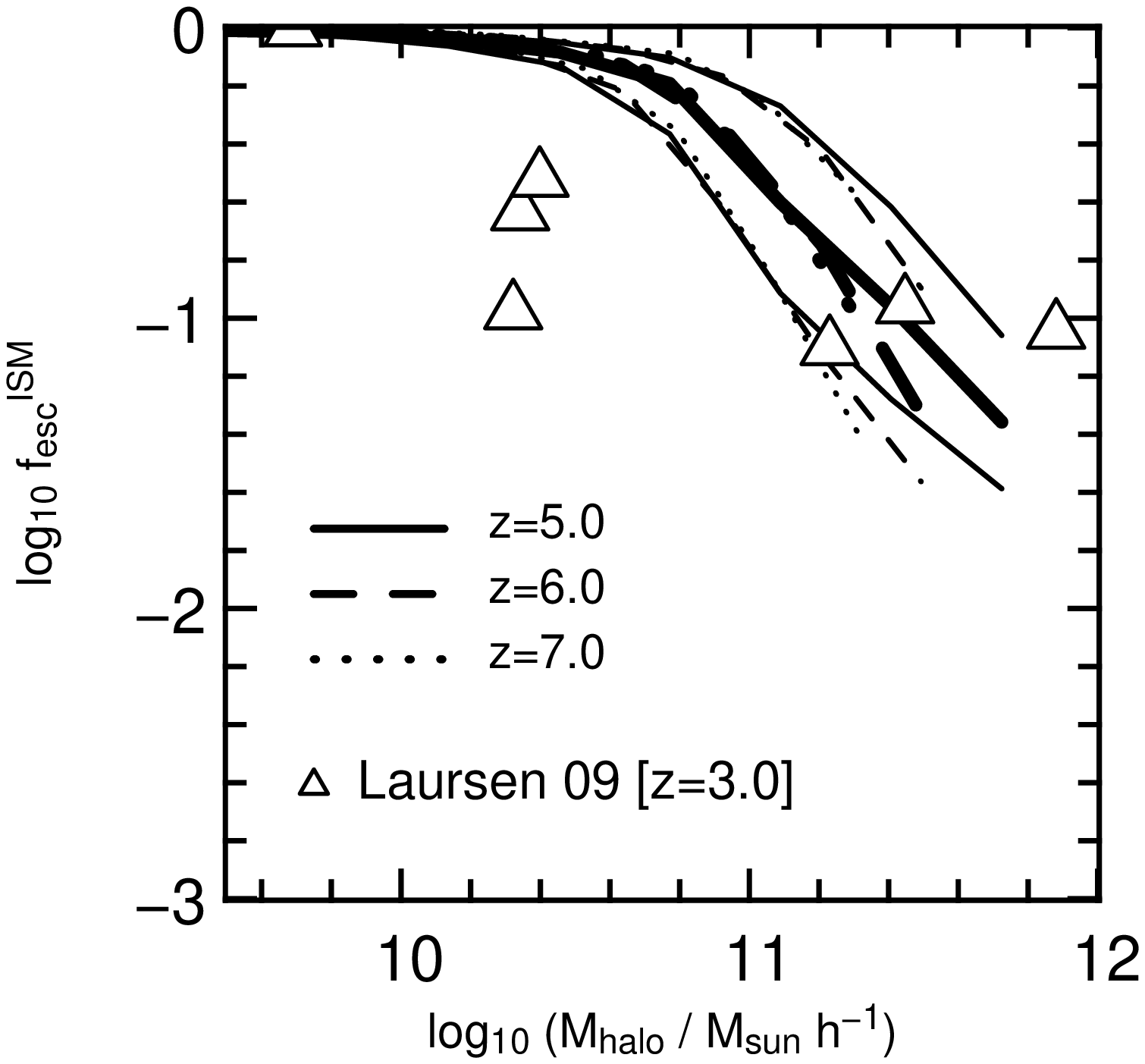}\hspace{0.5cm}
\includegraphics[width=0.45\textwidth]{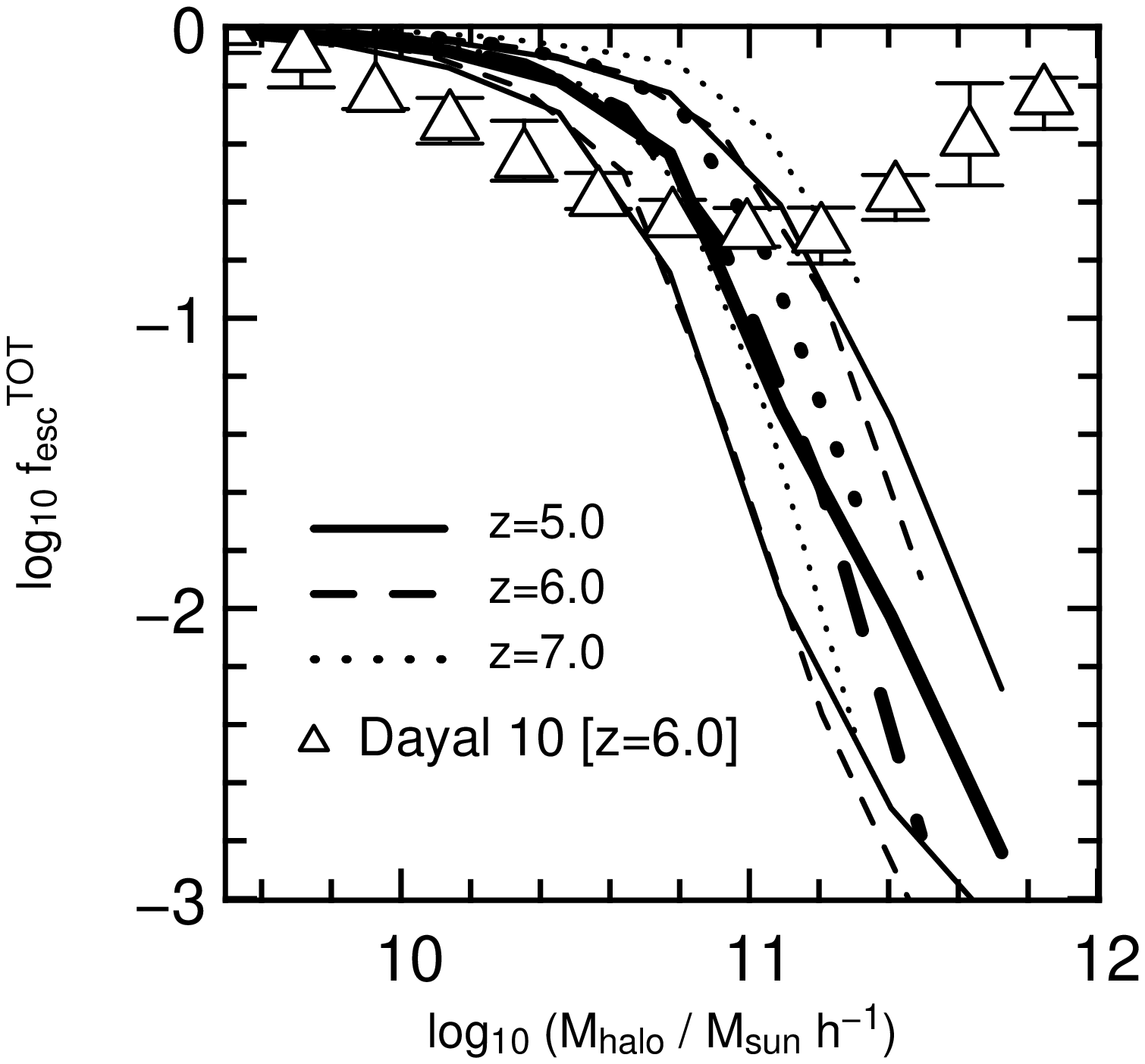}
\end{center}
\caption{Escape fraction as a function of dark matter halo mass, for   redshifts
  $z\sim 5$, $6$ and $7$. The thick lines  mark  the median,   and the thin
  lines represent the first and third quartile.  Triangles show different
  theoretical results. \emph{Left panel}. The escape fraction includes the
  contribution from the homogeneous ISM ($f_{esc}^{ISM}$) only. The median
  values for $f_{esc}^{ISM}$ are reproduced by Eq. (\ref{eq:f_esc_ism})  The
  symbols correspond to the results for single halo simulations by
  \citet{2009ApJ...704.1640L}.  \emph{Right panel}. Contribution from the
  birth clouds around the younger population stars to the escape fraction  ($f_{esc}^{BC}$).  
  The median values for $f_{esc}^{BC}$  are   reproduced by
  Eq.(\ref{eq:f_esc_bc}). The symbols    represent the
  results obtained by \citet{2010MNRAS.402.1449D} from a cosmological
  simulation.}
\label{fig:lya_escape_dm}
\end{figure*}

In the right panel of Figure \ref{fig:lya_escape_dm} we show the total
escape fraction $f_{esc}^{TOT}=f_{esc}^{ISM} \times f_{esc}^{BC}$ as a
function of the host halo mass for the homogeneous slab model, where
$f_{esc}^{ISM}$ and $f_{esc}^{BC}$ were defined in sub-section
\ref{sub:implement}. The overall scaling of $f_{esc}^{ISM}$ and
$f_{esc}^{BC}$ with halo mass follows the relation parameterized by
Eq.(\ref{eq:f_esc_homo}) where the product $\tau_{a}(a\tau_{0})^{1/3}$
seems to be proportional to the halo mass.

Due to this shape, the escape fraction has a large scatter for
galaxies in haloes more massive than $M_{c} = 6 \times 10^{10}\hMsun$.
The escape fraction for haloes less massive than this characteristic
mass $M_{c}$ seems to be always larger than $10\%$. 

We find that the median value of the escape fraction for the
homogeneous $ISM$ component, $f_{esc}^{ISM}$, between redshifts
$5\lesssim z \lesssim 7$ can be well approximated by the following
expression:   

\begin{equation}
  f_{esc}^{ISM} = \frac{1-e^{-R}}{R}, 
\label{eq:f_esc_ism}
\end{equation}
where $R$ depends on the halo mass, $M_h$, as follows:

\begin{equation}
R = \left(\frac{M_h}{M_c}\right)^{5/3}, 
\end{equation}
and the critical mass is equal to $M_c= 6\times 10^{10}\hMsun$. 

Correspondingly, the median value of the escape fraction associated
with the birth clouds of younger stellar populations, 
$f_{esc}^{BC}$, between redshifts $5\lesssim z\lesssim 7$ can be 
approximated by the following expression:    

\begin{equation}
f_{esc}^{BC}  = \frac{1 - e^{-D}}{D}, 
\label{eq:f_esc_bc}
\end{equation}
where $D$ is equal to 
\begin{equation}
D = \frac{1}{20}\left(\frac{1}{\mu} - 1\right)\left(\frac{M_{h}}{M_{c}}\right)^{5/3}
\end{equation}
and $M_{c}$ has the same value  as  in the case of the homogeneous ISM.
The results in the last two equations are dependent on the dust
abundances derived by matching the observed UV luminosity functions and
the UV colours in the way presented in \cite{2010MNRAS.403L..31F}. 

The observed scatter in $f_{esc}^{TOT}$ can be reproduced in a Monte
Carlo fashion if one calculates individual escape fractions
$f_{esc}^{ISM}$ and $f_{esc}^{BC}$ from  halo masses in the simulation,
using  Eqs. (\ref{eq:f_esc_ism}) and (\ref{eq:f_esc_bc}) with different
values of $M_{c}$ and taking $\log_{10} M_c/\hMsun$ as a Gaussian
variable with mean $\log_{10}(6\times10^{10})=10.77$ and variance of
$0.3$.

In  Figure \ref{fig:lya_escape_dm} we compare our results against two
other related theoretical works at high redshift. We start by pointing
out that the observed scaling in our results follows closely the
radiative transfer results shown in Figure \ref{fig:escape_sphere}. The
correlation  between halo mass and gaseous mass obtained in the
simulation  is translated here as a scaling between escape fraction and
halo mass.    

In the left panel of Figure \ref{fig:lya_escape_dm} we compare our results of the
escape fraction from the ISM component against the results of high resolution
simulations of individual galaxies \citep{2009ApJ...704.1640L}. The two
models show a good agreement for  high and low halo masses. There
is, however, a difference for haloes in  the  mass range  $\sim
3\times10^{10}$\hMsun. The mean of our escape fractions is higher by a factor of $\sim
4$, although the small sample of \cite{2009ApJ...704.1640L} (seven galaxies)
makes it difficult to estimate the statistical significance of this comparison
between the two models.

In the right panel of Figure \ref{fig:lya_escape_dm} we compare  our
results against a published model of the \lya escape fraction in a
cosmological context at $z\sim6$ \citep{2010MNRAS.402.1449D}.  The model
of \cite{2010MNRAS.402.1449D} presents a radically different prediction
for the escape fraction in haloes more massive than $M_{th} \sim
10^{11}$\hMsun. While in our model the mean escape fraction drops
steeply from $\sim 0.1$ down to $\sim 10^{-3}$, the results of
\cite{2010MNRAS.402.1449D} show an increase of the escape fraction to
values close to $\sim 1$ for the same range of massive haloes. This
disagreement is very likely due to the fact that
\cite{2010MNRAS.402.1449D} do not model explicitly the resonant nature
of the line and approximate the ISM in such a way that the \lya\ escape
fraction is proportional to $(1-\exp-(\tau_{d}))/\tau_{d}$ where
$\tau_{d}$ is the dust optical depth. This is a different approximation
from the one that produces the results represented by
Eq.\ref{eq:f_esc_homo} and other similar radiative transfer studies
\citep{2006MNRAS.367..979H} where there is an explicit dependence with
the optical depth of neutral gas, thought to be abundant in the most
massive and vigorous star forming galaxies at these redshifts.

\subsection{Intrinsic Emission and Luminosity Functions}
\label{subsection:intrinsic}

We now study the  LAE  LF predicted from our simulations taking into account
the different contributions to the escape fraction. We will focus on
the  intrinsic emission associated with star formation. The mechanism for
star-triggered \lya\ relates the amount of ionizing photons produced by young
stars to the expected intrinsic \lya\ luminosity.  Here we assume that the number of Hydrogen
ionizing photons per unit time is $1.8\times 10^{53}$ photons s$^{-1}$ for a
star formation rate of $1$ \Msun/yr \citep{1999ApJS..123....3L}. This value
assumes a Salpeter IMF, and that the star formation rate has been constant at
least during the last $10$ Myr. Assuming that $2/3$ of
these photons are converted to \lya photons (case-B recombination,
\citealt{1989agna.book.....O}), the intrinsic  \lya luminosity as a function of
the star formation rate is
\begin{equation}
L_{Ly\alpha} = 1.9 \times 10^{42}\times ({\mathrm{SFR}}/{\mathrm
  M}_{\odot}\ {\mathrm{yr}}^{-1}) {\ \mathrm{erg\ s}}^{-1}, 
\label{eq:lya_sfr}
\end{equation}
where the  SFR is calculated from the total mass of stars produced in the
last $200$ Myr.  We use this analytic expression because the
calculation of the total amount of ionizing photons directly from the
information of the star particles in the simulation is not reliable
given the short lifetime of the populations contributing to the ionizing flux and the limited
resolution of our simulation. Only the star particles younger than
$10$ Myr will provide the bulk of the ionizing flux, these particles
roughly correspond to $1\%$ to $5\%$ of the star particles in a given
galaxy. The least resolved galaxy in our study has $200$ star
particles, which correspond to $2$ to $10$ particles to fully sample
the star formation history of the galaxy during the last $10$
Myr. Only a handful of largest galaxies in the simulation ($\sim
100000$ star particles, $\sim 1000$ of them contributing to the
ionizing flux) can give reliable results when the \lya emission from
the star particles is compared against Eq. (\ref{eq:lya_sfr}).  

We find a tight correlation between star formation rate and halo mass
for $5\lesssim z\lesssim 7$. It can be approximated by
\begin{equation}
  {\mathrm{SFR}} = 0.68 \left(\frac{M_h}{10^{10} \hMsun}\right)^{1.30}
  M_{\odot}\ {\mathrm {yr}}^{-1},
\label{eq:sfr_hmass}
\end{equation}
so that the final scaling between intrinsic \lya emission and halo mass can be
written as
\begin{equation}
  L_{Ly\alpha} = 1.29 \times 10^{42} \left(\frac{M_h}{10^{10} \hMsun}\right)^{1.30} {\mathrm {erg\ s}}^{-1}.
\label{eq:lya_hmass}
\end{equation}

The observed \lya\ luminosity  can be calculated from our
  extinction model as:

\begin{equation}
L_{Ly\alpha}^{o} = (f^{ISM}_{esc} \times L_{Ly\alpha}^{old}) +  (f^{ISM}_{esc} \times f^{BC}_{esc} \times L_{Ly\alpha}^{young}),
\end{equation}
where the label \emph{old} refers to the \lya\ emission coming from
stellar populations older than $25$Myr and \emph{young} labels  the emission from
 stars younger  than $25$ Myr. Given that the intrinsic
ionizing flux is negligible for stellar  populations older than
$25$Myr, we can approximate  the observed \lya\ luminosity as:

\begin{equation}
L_{Ly\alpha}^{o} \approx f^{ISM}_{esc} \times f^{BC}_{esc} \times L_{Ly\alpha},
\label{eq:lya_obs}
\end{equation}
where  the escape fractions $f^{ISM}_{esc}$ and $f^{BC}_{esc}$  are calculated as
described in the previous subsections. 

Previous Monte-Carlo calculations of \lya radiative transfer in a
multiphase medium \citep{2006MNRAS.367..979H} show that the dominant
effect of the clumpy distribution of birth clouds on a photon
propagating through the homogeneous ISM is bouncing at the cloud's
surface, justifying the approximation of not including further
absorption by these clouds except at the \lya source in the way we have
just implemented.

However, we note that the conditions for dust and gas abundance in our
model are such that the calculations of \cite{1991ApJ...370L..85N} for
the escape fraction are not applicable here. The main assumption of that
model (negligible absorption and scattering in the homogeneous part of
the ISM) are not met, meaning that the escape fraction in our model is
not dominated by the effects of the clumpy component only.

We have not taken into account the effects of scattering of \lya
photons  by the IGM, after they escaped from the galaxy. This effect is
very important before reionisation ($z>6$ in our simulation).  After
reionisation,  it is expected that the blue part of the double peak
spectrum would be   absorbed as photons are cosmologically redshifted to
resonance with neutral Hydrogen along the line of sight. To first order,
we have taken this effect into account by dividing the intensity of the
observed line strength $L_{Ly\alpha}^{o}$ by $2$.

\begin{figure*}
\begin{center}
\includegraphics[width=0.33\textwidth]{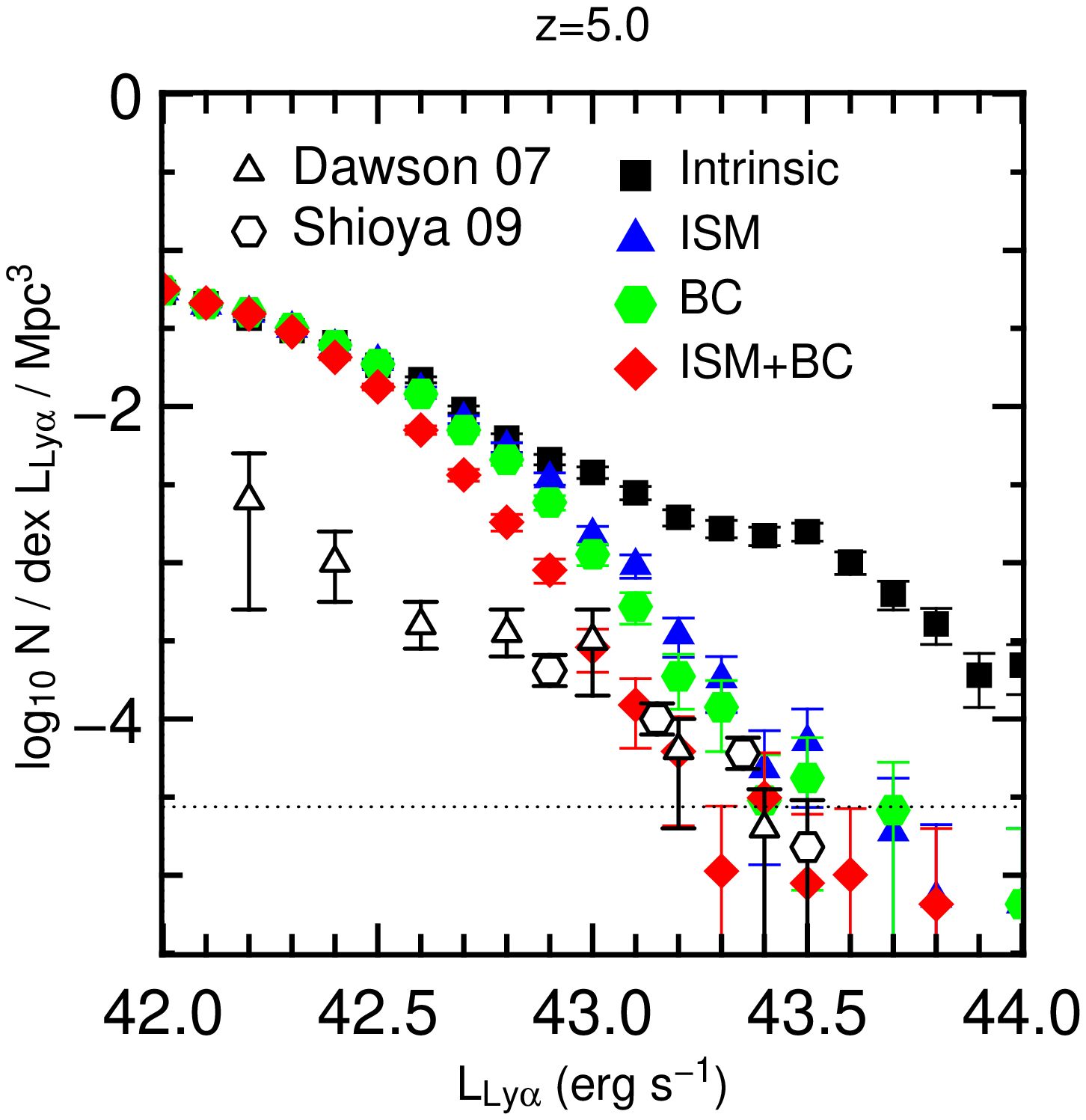}
\includegraphics[width=0.33\textwidth]{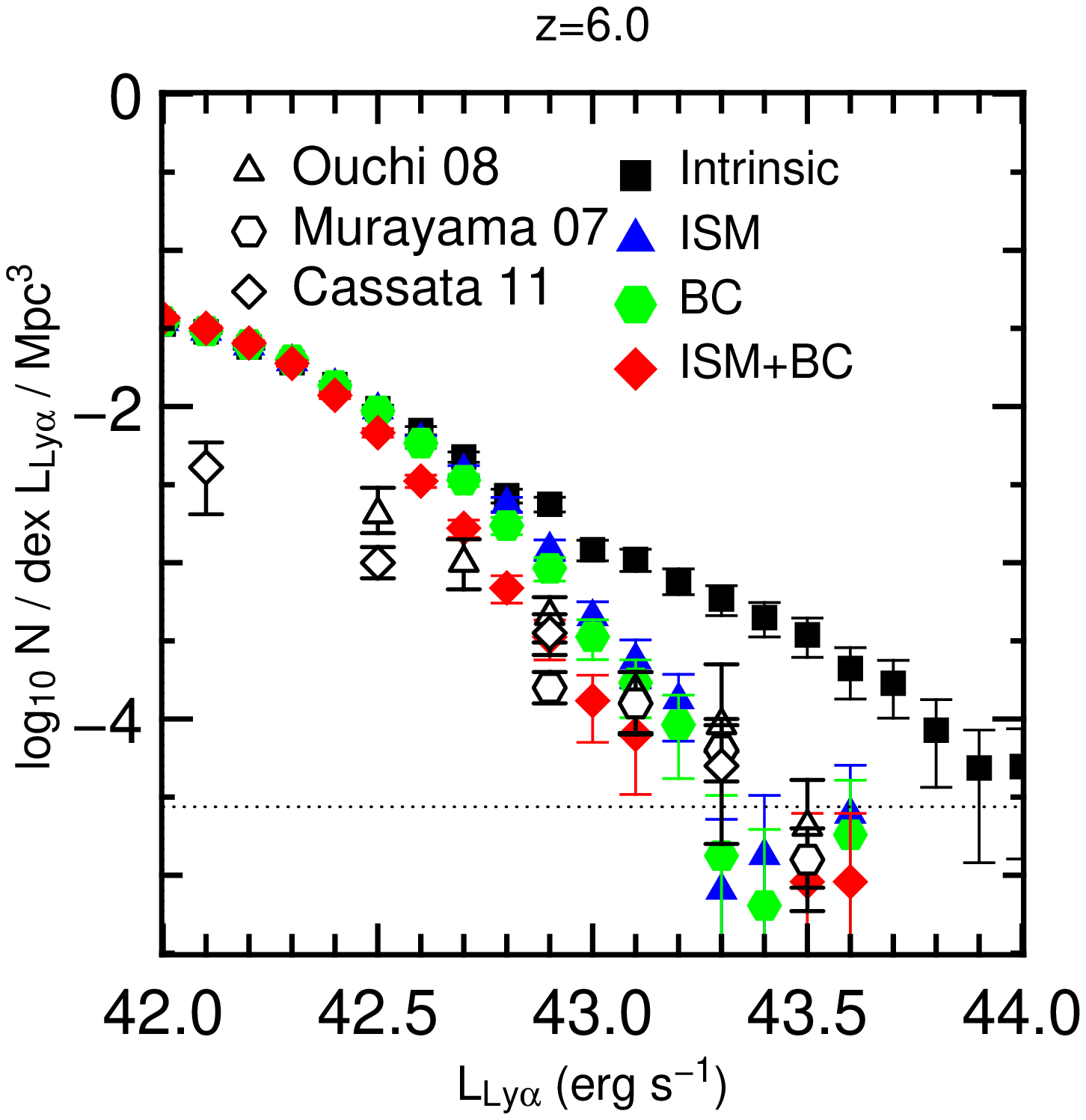}
\includegraphics[width=0.33\textwidth]{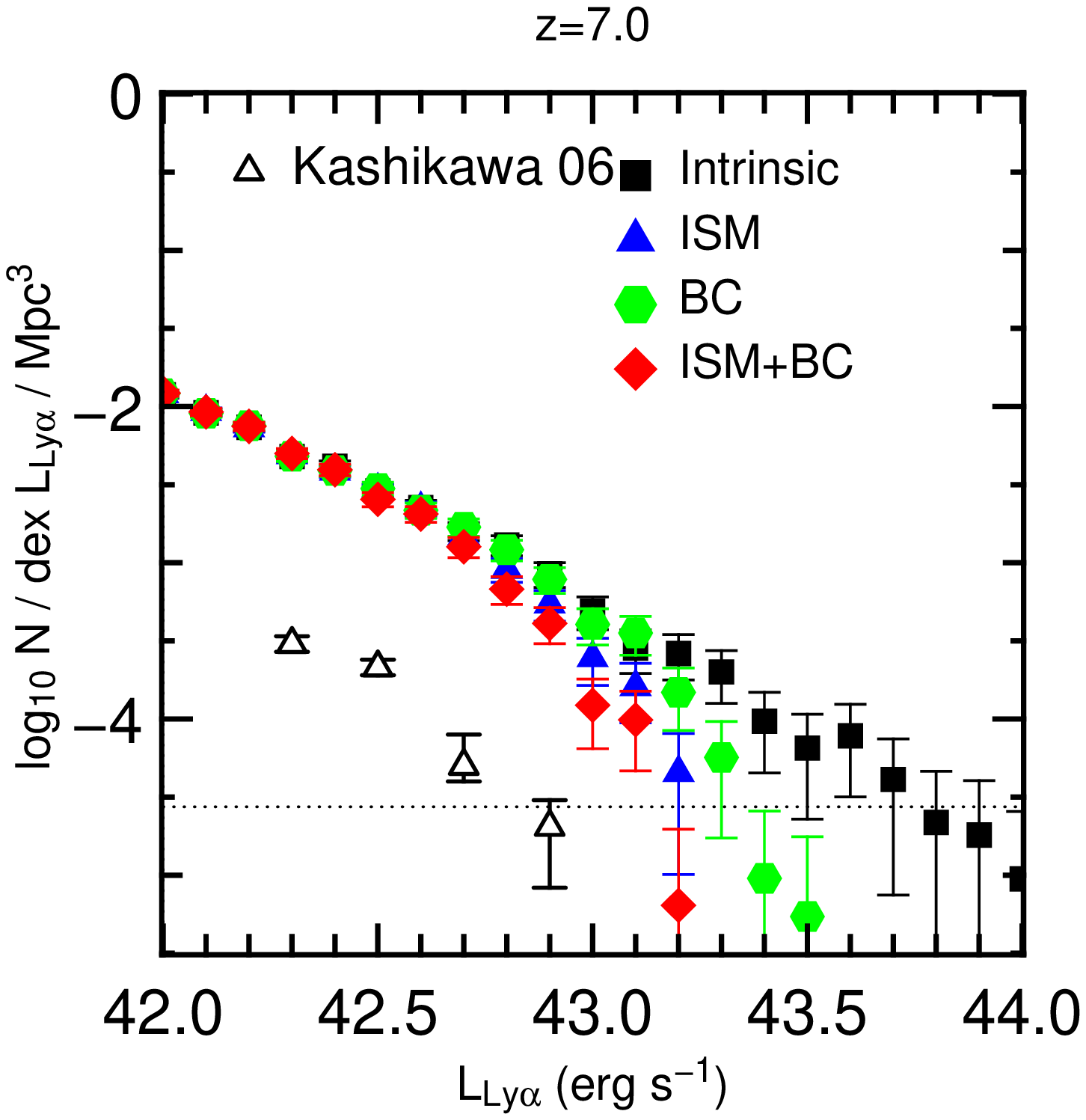}
\end{center}
\caption{Luminosity Functions at $z\sim 5,6$ and $7$. The observational constraints
  (empty symbols) are compared to different results from the simulation
  (filled symbols): Intrinsic \lya emission considering the total star
  formation rate in Eq.(\ref{eq:lya_sfr}) (black squares), \lya emission
  corrected only by the extinction of the homogeneous ISM based on the spherical
  model studied with the Monte Carlo simulations using Eq.(\ref{eq:f_esc_ism})
   (blue triangles), \lya emission corrected only  by the extinction of
  the birth clouds using Eq.(\ref{eq:f_esc_bc}) (green hexagons). The final luminosity function
  corrected by the homogeneous ISM and the birth clouds is shown by the red diamonds.}
\label{fig:lya_lf}
\end{figure*}

In Figure \ref{fig:lya_lf} we show three different LAE LFs at three different
redshifts $z\sim 5,6$ and $7$. The empty symbols represent the
observed LFs. The observational results at $z\sim 4.5$ come from 50
spectroscopically confirmed LAEs from the Large Area \lya survey
(LALA) which covered a field of $\sim 0.7$ deg$^2$ corresponding to a
comoving survey volume of $7.4\times 10^5$ Mpc$^3$
\citep{2007ApJ...671.1227D}. At $z\sim 4.86$ the results from
\cite{2009ApJ...696..546S} are based on observations of the Cosmic
Evolution Survey (COSMOS) field of $1.83$ deg$^2$ giving an effective volume
of $1.1\times 10^6$ Mpc$^{3}$. At $z\sim 6$ the data are taken from
\cite{2008ApJS..176..301O}, in this case the LF was estimated from
LAEs from 1 deg$^2$ Subaru/XMM-Newton Deep Survey (SXDS), which probed
a comoving volume of $\sim 10^6$ Mpc$^3$. We include as well data at $z\sim
5.7$ in a field of $1.95$ deg$^2$ covered by the COSMOS survey
\cite{2007ApJS..172..523M} and recent estimations at $4.5<z<6.6$ from
the Vimos-VLT Deep Survey (VVDS) \citep{2011A&A...525A.143C}.
The LF at $z\sim 7$ was constructed from a sample of 17 LAEs confirmed
spectroscopically and 58 photometric LAEs probing a comoving volume of
$\sim 2.17\times 10^5$ Mpc$^3$ \citep{2006ApJ...648....7K}. In our
simulation we probe a comoving volume that is of the same order of
magnitude of these surveys.

In Figure \ref{fig:lya_lf} we represent with black squares the LF
estimated from the intrinsic luminosities calculated from the star
formation rates. The spatial abundance is already corrected from the
different halo abundance between the WMAP5 and the WMAP1
cosmologies. Compared with the observational estimates at all
redshifts, the normalisation is at least one order of magnitude
higher.    

We can now apply the expected correction by the estimated escape fraction of
each galaxy. In our model there is a strong correlation between the
mass of the halo hosting the galaxy and its associated escape
fraction. In a simplified manner, galaxies in haloes with masses
smaller than $6.0 \times 10^{10}$ \hMsun remain practically
unaffected. This is seen in the LFs in Figure \ref{fig:lya_lf} where
the hexagons represent the luminosity function  corrected by the
escape fraction. The  bright end of the LF is modified but the overall
normalisation is kept one order of magnitude higher than observations.

This result  is completely analogous to the situation found in the continuum UV
using the same simulation \citep{2010MNRAS.403L..31F}. The strong scaling of
the reddening with galaxy mass causes only the bright end of the UV LF to be
effectively modified when extinction is taken into account. In the context of
our model, applying a constant reddening value to all galaxies can only be
physically justified if there is an additional extinction term from the
youngest stars, as described in Section \ref{sec:spectra}.

If we add now the expected correction of the birth clouds around stellar
populations younger than $25$ Myr, we find that the faint end of the LAE
LF can be further modified.  However, at $z\sim 5$ and $z\sim 6$ there
remains an excess at the faint end of the LF, corresponding to haloes
with masses $1.0\times10^{10}\hMsun < M_h < 4.0\times10^{10} \hMsun$.
This result suggests that either the escape fraction for these galaxies
or their star formation rate is too high. Nevertheless, the bright end
of the simulated LFs, both at $z\sim 5$ and $z\sim 6$, shows a very
close agreement with the observed one. The results are within Poissonian
uncertainty.  We have also checked that if we apply the scalings for the
intrinsic emission and escape fractions with halo mass, Eqs.
(\ref{eq:f_esc_ism}, \ref{eq:f_esc_bc}, \ref{eq:lya_hmass}), to a pure
DM only simulation with a cubic volume of $250$ \hMpc on a side and
$2048^3$ particles (the Bolshoi Simulation presented by
\cite{2010arXiv1002.3660K}) split into smaller sub-boxes, we  can
reproduce the results we have just discussed. Specifically, the scatter
at the bright end due to cosmic variance is consistent with the
observations. Nevertheless, the same scatter does not  help to explain
the  lower  abundance of  our numerical LAEs at the faint end.

The normalization of our LF functions  at redshift $z\sim 7$ is still
higher than observed. In principle, it  could be   possible  to account
for this difference by properly modeling the IGM absorption, which at
this epoch,  it  is not completely ionized and will add a dimming effect
to the  LAEs.  Considering a dimming factor  that depends on  luminosity
only (without any scatter), we would require that 25\% (40\%) of Lya
photons be  transmitted through the IGM  in order to  match the bright
(faint) end of the observed luminosity function. Using again the  DM
only simulation,  together with the scalings we obtain for the star
formation and the escape fraction, we find that cosmic variance can
account for a 0.5 dex (0.3 dex) variations at the bright (faint) end of
the luminosity function at $z\sim7$.  In conclusion, applying a constant
IGM transmission of $T=0.3-0.4$ we can still reproduce the luminosity
function for the brightest three bins in Figure \ref{fig:lya_lf}.  A
more realistic treatment of the effects of IGM (in the spirit of
\cite{2010ApJ...716..574Z}) actually yields a large scatter for the
transmission at a given \lya\ luminosity. Nevertheless, a detailed
modeling of the IGM effect is far beyond the scope of this  paper.

\section{Conclusions}
\label{sec:conclusions}

In this paper we model the escape fraction of \lya\ photons in the
approximation of a dusty gas slab with \lya\ sources homogeneously
mixed. The escape fractions for this configuration and different dust
and gas contents have been calculated using {\texttt{CLARA}}, our new
Monte-Carlo code described in detail in Appendix \ref{sec:montecarlo}.
These results can be applied in any model that predicts the optical
depth of gas and dust in galaxies.

We selected the slab geometrical configuration in order to be consistent
with the assumptions that led us to fix the dust abundances from the
\emph{MareNostrum   High-z Universe} simulation, as constrained by high
redshift UV at $z\gtrsim 5$ observations \citep{2010MNRAS.403L..31F}.
Our proposed dust model describes the contributions from an homogeneous
ISM (slab geometry) and a clumpy phase (spherical geometry) associated
to stellar populations younger than $25$Myr.

We estimate the scaling of the \lya\ escape fraction with the expected
reddening for the slab component and dust abundances in the
\emph{MareNostrum High-z Universe}. The scaling shows a weak redshift
dependence between $5\lesssim z \lesssim 7$, furthermore there is a very
good agreement with the observational estimation of the escape fraction
and reddening for galaxies at $z\sim 2.2$ and $z\sim 3$ assuming that
only the homogeneous ISM component is dominant at these redshifts.
Including both contributions from the homogeneous ISM and the clumpy
phase, we have calibrated the escape fraction as a function of the host
dark matter halo based on the results of the \emph{MaraNostrum High-z
Universe}.

As an application of these results, we construct the intrinsic LAEs LF
as estimated from the star formation rates. We find that the
normalisation is one order of magnitude higher than observational
estimates. Correcting the intrinsic LAE luminosities by the estimated
escape fraction (homogeneous ISM and clumpy phase included) brings the
simulation into agreement with the observations  at $z\sim 5$ and $z\sim
6$. The match at the bright end is acceptable within the Poissonian and
cosmic variance errors. The mismatch at $z\sim 7$ can be explained
because a proper modeling of these epochs has to account for the yet
incomplete reionisation process and the influence of the neutral parts
of the IGM.

Nevertheless,  our results are in conflict with the observational
estimates at the faint end. There seems to be an excessive production of
LAEs with intrinsic luminosities $L_{L\alpha}^{o}\sim 6.0\times 10^{42}$
erg s$^{-1}$ at $z\sim5$ and $z\sim6$.  This excess was spotted,
although weakly, in the results of the UV LFs  of
\citep{2010MNRAS.403L..31F}.  Furthermore, a  fine tuning of the
extinction model for galaxies in this mass range could provide a better
match to the observations. In that case, the galaxies at  the faint end
should be more dusty, which would make them  redder, and thus breaking
the broad agreement for the UV colours that we have already obtained.
This is not a satisfactory approach to explain  for these  differences.
Based on  these considerations,  we think that a probable origin of this
discrepancy is the high rate of star formation in galaxies situated in
these haloes. A possible physical explanation is that supernova feedback
modulates more effectively the star formation in haloes of masses
$<10^{11}$\hMsun, providing a mechanism to shape the faint end of the
luminosity function. However, it is also possible that the star
formation rate is overestimated in all haloes at these redshifts. This
would be translated into a trade-off with less dust extinction to
explain the UV luminosity function. Regardless of what is the correct
explanation of this enigma, all possible solutions seem to challenge our
current understanding of the rate  at which gas is converted into stars
at high redshift.

It is encouraging that the results for the brightest galaxies, the best
resolved ones,  both in observations and simulations, are consistent
with observations in the UV (magnitudes and colours) as well with the
Lyman-$\alpha$ line. A full picture of these massive high redshift
galaxies will be completed by observations of the rest frame IR to be
performed by ALMA. We will address in a upcoming work the predictions of
our model in terms of ALMA observations, focusing on the most massive
galaxies, constructing a complete panchromatic perspective of high
redshift galaxies in the {\it MareNostrum High-z  Universe}.

\section*{Acknowledgments}

The simulation used in this work is part of the MareNostrum Numerical
Cosmology Project at the BSC. The data analysis has been performed at
the NIC J\"ulich and at the LRZ Munich.

The Bolshoi simulation used in this paper was performed and analyzed at
the NASA Ames Research Center. We thank A. Klypin (NMSU) and J. Primack
(UCSC) for making this  simulation available to us.

JEFR and FP acknowledge the support by the ESF ASTROSIM network though a
short visit grant of JEFR to Granada where part of the developing and
most of the testing  for \clara took place. JEFR acknowledges the
hospitality of Roberto Luccas in Barcelona, where most of this paper was
written. JEFR acknowledges as well useful discussions on some issues
addressed in this paper with Renyue Cen, Zheng Zheng and Chung-Pei Ma.
JEFR thanks Pratika Dayal for providing data from her paper in
electronic format.

GY acknowledges support of  MICINN  (Spain) through research grants
FPA2009-08958 and  AYA2009-13875-C03-02. SRK would like to thank
Consolider-Ingenio SyeC (Spain)  (CSD2007-0050) for financial support. 

AJC acknowledges support from MICINN through FPU grant 2005-1826.

We equally acknowledge funding from the  Consolider  project MULTIDARK
(CSD2009-00064) and the Comunidad de Madrid project ASTROMADRID
(S2009/ESP-146).

\bibliographystyle{mn2e}

\begin{thebibliography}{}

\bibitem[\protect\citeauthoryear{{Calzetti}, {Armus}, {Bohlin}, {Kinney},
  {Koornneef} \& {Storchi-Bergmann}}{{Calzetti}
  et~al.}{2000}]{2000ApJ...533..682C}
{Calzetti} D.,  {Armus} L.,  {Bohlin} R.~C.,  {Kinney} A.~L.,  {Koornneef} J.,
    {Storchi-Bergmann} T.,  2000, \apj, 533, 682

\bibitem[\protect\citeauthoryear{{Cassata}, {Le F{\`e}vre}, {Garilli},
  {Maccagni}, {Le Brun}, {Scodeggio}, {Tresse} \& {Ilbert} O.~{et
  al}}{{Cassata} et~al.}{2011}]{2011A&A...525A.143C}
{Cassata} P.,  {Le F{\`e}vre} O.,  {Garilli} B.,  {Maccagni} D.,  {Le Brun} V.,
   {Scodeggio} M.,  {Tresse} L.,    {Ilbert} O.~{et al} .,  2011, \aap, 525,
  A143+

\bibitem[\protect\citeauthoryear{{Cazaux} \& {Spaans}}{{Cazaux} \&
  {Spaans}}{2004}]{2004ApJ...611...40C}
{Cazaux} S.,  {Spaans} M.,  2004, \apj, 611, 40

\bibitem[\protect\citeauthoryear{{Ceverino}, {Dekel} \& {Bournaud}}{{Ceverino}
  et~al.}{2010}]{2010MNRAS.404.2151C}
{Ceverino} D.,  {Dekel} A.,    {Bournaud} F.,  2010, \mnras, 404, 2151

\bibitem[\protect\citeauthoryear{{Charlot} \& {Fall}}{{Charlot} \&
  {Fall}}{2000}]{2000ApJ...539..718C}
{Charlot} S.,  {Fall} S.~M.,  2000, \apj, 539, 718

\bibitem[\protect\citeauthoryear{{Dawson}, {Rhoads}, {Malhotra}, {Stern},
  {Wang}, {Dey}, {Spinrad} \& {Jannuzi}}{{Dawson}
  et~al.}{2007}]{2007ApJ...671.1227D}
{Dawson} S.,  {Rhoads} J.~E.,  {Malhotra} S.,  {Stern} D.,  {Wang} J.,  {Dey}
  A.,  {Spinrad} H.,    {Jannuzi} B.~T.,  2007, \apj, 671, 1227

\bibitem[\protect\citeauthoryear{{Dayal}, {Ferrara} \& {Saro}}{{Dayal}
  et~al.}{2010}]{2010MNRAS.402.1449D}
{Dayal} P.,  {Ferrara} A.,    {Saro} A.,  2010, \mnras, 402, 1449

\bibitem[\protect\citeauthoryear{{Devriendt}, {Rimes}, {Pichon}, {Teyssier},
  {Le Borgne}, {Aubert}, {Audit}, {Colombi}, {Courty}, {Dubois}, {Prunet},
  {Rasera}, {Slyz} \& {Tweed}}{{Devriendt} et~al.}{2010}]{2010MNRAS.403L..84D}
{Devriendt} J.,  {Rimes} C.,  {Pichon} C.,  {Teyssier} R.,  {Le Borgne} D.,
  {Aubert} D.,  {Audit} E.,  {Colombi} S.,  {Courty} S.,  {Dubois} Y.,
  {Prunet} S.,  {Rasera} Y.,  {Slyz} A.,    {Tweed} D.,  2010, \mnras, 403, L84

\bibitem[\protect\citeauthoryear{{Devriendt}, {Guiderdoni} \&
  {Sadat}}{{Devriendt} et~al.}{1999}]{1999A&A...350..381D}
{Devriendt} J.~E.~G.,  {Guiderdoni} B.,    {Sadat} R.,  1999, \aap, 350, 381

\bibitem[\protect\citeauthoryear{{Dijkstra}, {Haiman} \& {Spaans}}{{Dijkstra}
  et~al.}{2006}]{2006ApJ...649...14D}
{Dijkstra} M.,  {Haiman} Z.,    {Spaans} M.,  2006, \apj, 649, 14

\bibitem[\protect\citeauthoryear{{Dunkley}, {Komatsu}, {Nolta}, {Spergel},
  {Larson}, {Hinshaw}, {Page}, {Bennett}, {Gold}, {Jarosik}, {Weiland},
  {Halpern}, {Hill}, {Kogut}, {Limon}, {Meyer}, {Tucker}, {Wollack} \&
  {Wright}}{{Dunkley} et~al.}{2009}]{2009ApJS..180..306D}
{Dunkley} J.,  {Komatsu} E.,  {Nolta} M.~R.,  {Spergel} D.~N.,  {Larson} D.,
  {Hinshaw} G.,  {Page} L.,  {Bennett} C.~L.,  {Gold} B.,  {Jarosik} N.,
  {Weiland} J.~L.,  {Halpern} M.,  {Hill} R.~S.,  {Kogut} A.,  {Limon} M.,
  {Meyer} S.~S.,  {Tucker} G.~S.,  {Wollack} E.,    {Wright} E.~L.,  2009,
  \apjs, 180, 306

\bibitem[\protect\citeauthoryear{{Eisenstein et al.}}{{Eisenstein et
  al.}}{2005}]{2005ApJ...633..560E}
{Eisenstein et al.} 2005, \apj, 633, 560

\bibitem[\protect\citeauthoryear{{Ferri{\`e}re}}{{Ferri{\`e}re}}{2001}]{2001Rv%
MP...73.1031F}
{Ferri{\`e}re} K.~M.,  2001, Reviews of Modern Physics, 73, 1031

\bibitem[\protect\citeauthoryear{{Finlator}, {Oppenheimer} \&
  {Dav{\'e}}}{{Finlator} et~al.}{2011}]{2011MNRAS.410.1703F}
{Finlator} K.,  {Oppenheimer} B.~D.,    {Dav{\'e}} R.,  2011, \mnras, 410, 1703

\bibitem[\protect\citeauthoryear{{Forero-Romero}, {Yepes}, {Gottl{\"o}ber},
  {Knollmann}, {Khalatyan}, {Cuesta} \& {Prada}}{{Forero-Romero}
  et~al.}{2010}]{2010MNRAS.403L..31F}
{Forero-Romero} J.~E.,  {Yepes} G.,  {Gottl{\"o}ber} S.,  {Knollmann} S.~R.,
  {Khalatyan} A.,  {Cuesta} A.~J.,    {Prada} F.,  2010, \mnras, 403, L31

\bibitem[\protect\citeauthoryear{{Hansen} \& {Oh}}{{Hansen} \&
  {Oh}}{2006}]{2006MNRAS.367..979H}
{Hansen} M.,  {Oh} S.~P.,  2006, \mnras, 367, 979

\bibitem[\protect\citeauthoryear{{Harrington}}{{Harrington}}{1973}]{1973MNRAS.%
162...43H}
{Harrington} J.~P.,  1973, \mnras, 162, 43

\bibitem[\protect\citeauthoryear{{Hatton}, {Devriendt}, {Ninin}, {Bouchet},
  {Guiderdoni} \& {Vibert}}{{Hatton} et~al.}{2003}]{2003MNRAS.343...75H}
{Hatton} S.,  {Devriendt} J.~E.~G.,  {Ninin} S.,  {Bouchet} F.~R.,
  {Guiderdoni} B.,    {Vibert} D.,  2003, \mnras, 343, 75

\bibitem[\protect\citeauthoryear{{Hayes}, {{\"O}stlin}, {Schaerer},
  {Mas-Hesse}, {Leitherer}, {Atek}, {Kunth}, {Verhamme}, {de Barros} \&
  {Melinder}}{{Hayes} et~al.}{2010}]{2010Natur.464..562H}
{Hayes} M.,  {{\"O}stlin} G.,  {Schaerer} D.,  {Mas-Hesse} J.~M.,  {Leitherer}
  C.,  {Atek} H.,  {Kunth} D.,  {Verhamme} A.,  {de Barros} S.,    {Melinder}
  J.,  2010, \nat, 464, 562

\bibitem[\protect\citeauthoryear{{Hill}, {Gebhardt}, {Komatsu}, {Drory},
  {MacQueen}, {Adams}, {Blanc}, {Koehler}, {Rafal}, {Roth}, {Kelz}, {Gronwall},
  {Ciardullo} \& {Schneider}}{{Hill} et~al.}{2008}]{2008ASPC..399..115H}
{Hill} G.~J.,  {Gebhardt} K.,  {Komatsu} E.,  {Drory} N.,  {MacQueen} P.~J.,
  {Adams} J.,  {Blanc} G.~A.,  {Koehler} R.,  {Rafal} M.,  {Roth} M.~M.,
  {Kelz} A.,  {Gronwall} C.,  {Ciardullo} R.,    {Schneider} D.~P.,  2008, in
  {T.~Kodama, T.~Yamada, \& K.~Aoki} ed., Astronomical Society of the Pacific
  Conference Series Vol.~399 of Astronomical Society of the Pacific Conference
  Series, {The Hobby-Eberly Telescope Dark Energy Experiment (HETDEX):
  Description and Early Pilot Survey Results}.
pp 115--+

\bibitem[\protect\citeauthoryear{{Hirashita}, {Nozawa}, {Kozasa}, {Ishii} \&
  {Takeuchi}}{{Hirashita} et~al.}{2005}]{2005MNRAS.357.1077H}
{Hirashita} H.,  {Nozawa} T.,  {Kozasa} T.,  {Ishii} T.~T.,    {Takeuchi}
  T.~T.,  2005, \mnras, 357, 1077

\bibitem[\protect\citeauthoryear{{Hu} \& {McMahon}}{{Hu} \&
  {McMahon}}{1996}]{1996Natur.382..281H}
{Hu} E.,  {McMahon} R.~G.,  1996, \nat, 382, 281

\bibitem[\protect\citeauthoryear{{Hu}, {Cowie}, {Capak} \& {Kakazu}}{{Hu}
  et~al.}{2005}]{2005pgqa.conf..363H}
{Hu} E.~M.,  {Cowie} L.~L.,  {Capak} P.,    {Kakazu} Y.,  2005, in
  {P.~Williams, C.-G.~Shu, \& B.~Menard} ed., IAU Colloq. 199: Probing Galaxies
  through Quasar Absorption Lines {Spectroscopic studies of {$z \tilde 5.7$}
  and {$z \tilde 6.5$} galaxies: implications for reionisation}.
pp 363--368

\bibitem[\protect\citeauthoryear{{Hu}, {Cowie}, {Capak}, {McMahon}, {Hayashino}
  \& {Komiyama}}{{Hu} et~al.}{2004}]{2004AJ....127..563H}
{Hu} E.~M.,  {Cowie} L.~L.,  {Capak} P.,  {McMahon} R.~G.,  {Hayashino} T.,
  {Komiyama} Y.,  2004, \aj, 127, 563

\bibitem[\protect\citeauthoryear{{Hu}, {Cowie} \& {McMahon}}{{Hu}
  et~al.}{1998}]{1998ApJ...502L..99H}
{Hu} E.~M.,  {Cowie} L.~L.,    {McMahon} R.~G.,  1998, \apjl, 502, L99+

\bibitem[\protect\citeauthoryear{{Hu}, {Cowie}, {McMahon}, {Capak}, {Iwamuro},
  {Kneib}, {Maihara} \& {Motohara}}{{Hu} et~al.}{2002}]{2002ApJ...568L..75H}
{Hu} E.~M.,  {Cowie} L.~L.,  {McMahon} R.~G.,  {Capak} P.,  {Iwamuro} F.,
  {Kneib} J.,  {Maihara} T.,    {Motohara} K.,  2002, \apjl, 568, L75

\bibitem[\protect\citeauthoryear{{Inoue}}{{Inoue}}{2003}]{2003PASJ...55..901I}
{Inoue} A.~K.,  2003, \pasj, 55, 901

\bibitem[\protect\citeauthoryear{{Inoue}}{{Inoue}}{2005}]{2005MNRAS.359..171I}
{Inoue} A.~K.,  2005, \mnras, 359, 171

\bibitem[\protect\citeauthoryear{{Kashikawa}, {Shimasaku}, {Malkan}, {Doi},
  {Matsuda}, {Ouchi}, {Taniguchi}, {Ly}, {Nagao}, {Iye}, {Motohara},
  {Murayama}, {Murozono}, {Nariai}, {Ohta}, {Okamura}, {Sasaki}, {Shioya} \&
  {Umemura}}{{Kashikawa} et~al.}{2006}]{2006ApJ...648....7K}
{Kashikawa} N.,  {Shimasaku} K.,  {Malkan} M.~A.,  {Doi} M.,  {Matsuda} Y.,
  {Ouchi} M.,  {Taniguchi} Y.,  {Ly} C.,  {Nagao} T.,  {Iye} M.,  {Motohara}
  K.,  {Murayama} T.,  {Murozono} K.,  {Nariai} K.,  {Ohta} K.,  {Okamura} S.,
  {Sasaki} T.,  {Shioya} Y.,    {Umemura} M.,  2006, \apj, 648, 7

\bibitem[\protect\citeauthoryear{{Klypin}, {Trujillo-Gomez} \&
  {Primack}}{{Klypin} et~al.}{2010}]{2010arXiv1002.3660K}
{Klypin} A.,  {Trujillo-Gomez} S.,    {Primack} J.,  2010, ArXiv e-prints

\bibitem[\protect\citeauthoryear{{Knollmann} \& {Knebe}}{{Knollmann} \&
  {Knebe}}{2009}]{Knollmann2009}
{Knollmann} S.~R.,  {Knebe} A.,  2009, \apjs, 182, 608

\bibitem[\protect\citeauthoryear{{Kobayashi}, {Totani} \&
  {Nagashima}}{{Kobayashi} et~al.}{2007}]{2007ApJ...670..919K}
{Kobayashi} M.~A.~R.,  {Totani} T.,    {Nagashima} M.,  2007, \apj, 670, 919

\bibitem[\protect\citeauthoryear{{Kong}, {Charlot}, {Brinchmann} \&
  {Fall}}{{Kong} et~al.}{2004}]{2004MNRAS.349..769K}
{Kong} X.,  {Charlot} S.,  {Brinchmann} J.,    {Fall} S.~M.,  2004, \mnras,
  349, 769

\bibitem[\protect\citeauthoryear{{Kornei}, {Shapley}, {Erb}, {Steidel},
  {Reddy}, {Pettini} \& {Bogosavljevi{\'c}}}{{Kornei}
  et~al.}{2010}]{2010ApJ...711..693K}
{Kornei} K.~A.,  {Shapley} A.~E.,  {Erb} D.~K.,  {Steidel} C.~C.,  {Reddy}
  N.~A.,  {Pettini} M.,    {Bogosavljevi{\'c}} M.,  2010, \apj, 711, 693

\bibitem[\protect\citeauthoryear{{Laursen}, {Sommer-Larsen} \&
  {Andersen}}{{Laursen} et~al.}{2009}]{2009ApJ...704.1640L}
{Laursen} P.,  {Sommer-Larsen} J.,    {Andersen} A.~C.,  2009, \apj, 704, 1640

\bibitem[\protect\citeauthoryear{{Le Delliou}, {Lacey}, {Baugh}, {Guiderdoni},
  {Bacon}, {Courtois}, {Sousbie} \& {Morris}}{{Le Delliou}
  et~al.}{2005}]{2005MNRAS.357L..11L}
{Le Delliou} M.,  {Lacey} C.,  {Baugh} C.~M.,  {Guiderdoni} B.,  {Bacon} R.,
  {Courtois} H.,  {Sousbie} T.,    {Morris} S.~L.,  2005, \mnras, 357, L11

\bibitem[\protect\citeauthoryear{{Leitherer}, {Schaerer}, {Goldader},
  {Gonz{\'a}lez Delgado}, {Robert}, {Kune}, {de Mello}, {Devost} \&
  {Heckman}}{{Leitherer} et~al.}{1999}]{1999ApJS..123....3L}
{Leitherer} C.,  {Schaerer} D.,  {Goldader} J.~D.,  {Gonz{\'a}lez Delgado}
  R.~M.,  {Robert} C.,  {Kune} D.~F.,  {de Mello} D.~F.,  {Devost} D.,
  {Heckman} T.~M.,  1999, \apjs, 123, 3

\bibitem[\protect\citeauthoryear{{Malhotra} \& {Rhoads}}{{Malhotra} \&
  {Rhoads}}{2004}]{2004ApJ...617L...5M}
{Malhotra} S.,  {Rhoads} J.~E.,  2004, \apjl, 617, L5

\bibitem[\protect\citeauthoryear{{Mathis}, {Mezger} \& {Panagia}}{{Mathis}
  et~al.}{1983}]{1983A&A...128..212M}
{Mathis} J.~S.,  {Mezger} P.~G.,    {Panagia} N.,  1983, \aap, 128, 212

\bibitem[\protect\citeauthoryear{{Murayama}, {Taniguchi}, {Scoville}, {Ajiki},
  {Sanders}, {Mobasher}, {Aussel}, {Capak}, {Koekemoer}, {Shioya}, {Nagao},
  {Carilli}, {Ellis}, {Garilli} \& {Giavalisco}}{{Murayama}
  et~al.}{2007}]{2007ApJS..172..523M}
{Murayama} T.,  {Taniguchi} Y.,  {Scoville} N.~Z.,  {Ajiki} M.,  {Sanders}
  D.~B.,  {Mobasher} B.,  {Aussel} H.,  {Capak} P.,  {Koekemoer} A.,  {Shioya}
  Y.,  {Nagao} T.,  {Carilli} C.,  {Ellis} R.~S.,  {Garilli} B.,
  {Giavalisco} M.,  2007, \apjs, 172, 523

\bibitem[\protect\citeauthoryear{{Neufeld}}{{Neufeld}}{1991}]{1991ApJ...370L..%
85N}
{Neufeld} D.~A.,  1991, \apjl, 370, L85

\bibitem[\protect\citeauthoryear{{Night}, {Nagamine}, {Springel} \&
  {Hernquist}}{{Night} et~al.}{2006}]{2006MNRAS.366..705N}
{Night} C.,  {Nagamine} K.,  {Springel} V.,    {Hernquist} L.,  2006, \mnras,
  366, 705

\bibitem[\protect\citeauthoryear{{Nilsson}, {Orsi}, {Lacey}, {Baugh} \&
  {Thommes}}{{Nilsson} et~al.}{2007}]{2007A&A...474..385N}
{Nilsson} K.~K.,  {Orsi} A.,  {Lacey} C.~G.,  {Baugh} C.~M.,    {Thommes} E.,
  2007, \aap, 474, 385

\bibitem[\protect\citeauthoryear{{Osterbrock}}{{Osterbrock}}{1989}]{1989agna.b%
ook.....O}
{Osterbrock} D.~E.,  1989, {Astrophysics of gaseous nebulae and active galactic
  nuclei}

\bibitem[\protect\citeauthoryear{{Ota}, {Iye}, {Kashikawa}, {Shimasaku},
  {Kobayashi}, {Totani}, {Nagashima}, {Morokuma}, {Furusawa}, {Hattori},
  {Matsuda}, {Hashimoto} \& {Ouchi}}{{Ota} et~al.}{2008}]{2008ApJ...677...12O}
{Ota} K.,  {Iye} M.,  {Kashikawa} N.,  {Shimasaku} K.,  {Kobayashi} M.,
  {Totani} T.,  {Nagashima} M.,  {Morokuma} T.,  {Furusawa} H.,  {Hattori} T.,
  {Matsuda} Y.,  {Hashimoto} T.,    {Ouchi} M.,  2008, \apj, 677, 12

\bibitem[\protect\citeauthoryear{{Ouchi}, {Mobasher}, {Shimasaku}, {Ferguson},
  {Fall}, {Ono}, {Kashikawa}, {Morokuma}, {Nakajima}, {Okamura}, {Dickinson},
  {Giavalisco} \& {Ohta}}{{Ouchi} et~al.}{2009}]{2009ApJ...706.1136O}
{Ouchi} M.,  {Mobasher} B.,  {Shimasaku} K.,  {Ferguson} H.~C.,  {Fall} S.~M.,
  {Ono} Y.,  {Kashikawa} N.,  {Morokuma} T.,  {Nakajima} K.,  {Okamura} S.,
  {Dickinson} M.,  {Giavalisco} M.,    {Ohta} K.,  2009, \apj, 706, 1136

\bibitem[\protect\citeauthoryear{{Ouchi}, {Shimasaku}, {Akiyama}, {Simpson},
  {Saito}, {Ueda}, {Furusawa}, {Sekiguchi}, {Yamada}, {Kodama}, {Kashikawa},
  {Okamura}, {Iye}, {Takata}, {Yoshida} \& {Yoshida}}{{Ouchi}
  et~al.}{2008}]{2008ApJS..176..301O}
{Ouchi} M.,  {Shimasaku} K.,  {Akiyama} M.,  {Simpson} C.,  {Saito} T.,  {Ueda}
  Y.,  {Furusawa} H.,  {Sekiguchi} K.,  {Yamada} T.,  {Kodama} T.,  {Kashikawa}
  N.,  {Okamura} S.,  {Iye} M.,  {Takata} T.,  {Yoshida} M.,    {Yoshida} M.,
  2008, \apjs, 176, 301

\bibitem[\protect\citeauthoryear{{Partridge} \& {Peebles}}{{Partridge} \&
  {Peebles}}{1967}]{1967ApJ...147..868P}
{Partridge} R.~B.,  {Peebles} P.~J.~E.,  1967, \apj, 147, 868

\bibitem[\protect\citeauthoryear{{Reddy}, {Steidel}, {Fadda}, {Yan}, {Pettini},
  {Shapley}, {Erb} \& {Adelberger}}{{Reddy} et~al.}{2006}]{2006ApJ...644..792R}
{Reddy} N.~A.,  {Steidel} C.~C.,  {Fadda} D.,  {Yan} L.,  {Pettini} M.,
  {Shapley} A.~E.,  {Erb} D.~K.,    {Adelberger} K.~L.,  2006, \apj, 644, 792

\bibitem[\protect\citeauthoryear{{Rhoads}, {Dey}, {Malhotra}, {Stern},
  {Spinrad}, {Jannuzi}, {Dawson}, {Brown} \& {Landes}}{{Rhoads}
  et~al.}{2003}]{2003AJ....125.1006R}
{Rhoads} J.~E.,  {Dey} A.,  {Malhotra} S.,  {Stern} D.,  {Spinrad} H.,
  {Jannuzi} B.~T.,  {Dawson} S.,  {Brown} M.~J.~I.,    {Landes} E.,  2003, \aj,
  125, 1006

\bibitem[\protect\citeauthoryear{{Shapley}, {Steidel}, {Adelberger},
  {Dickinson}, {Giavalisco} \& {Pettini}}{{Shapley}
  et~al.}{2001}]{2001ApJ...562...95S}
{Shapley} A.~E.,  {Steidel} C.~C.,  {Adelberger} K.~L.,  {Dickinson} M.,
  {Giavalisco} M.,    {Pettini} M.,  2001, \apj, 562, 95

\bibitem[\protect\citeauthoryear{{Shimasaku}, {Kashikawa}, {Doi}, {Ly},
  {Malkan}, {Matsuda}, {Ouchi}, {Hayashino}, {Iye}, {Motohara}, {Murayama},
  {Nagao}, {Ohta}, {Okamura}, {Sasaki}, {Shioya} \& {Taniguchi}}{{Shimasaku}
  et~al.}{2006}]{2006PASJ...58..313S}
{Shimasaku} K.,  {Kashikawa} N.,  {Doi} M.,  {Ly} C.,  {Malkan} M.~A.,
  {Matsuda} Y.,  {Ouchi} M.,  {Hayashino} T.,  {Iye} M.,  {Motohara} K.,
  {Murayama} T.,  {Nagao} T.,  {Ohta} K.,  {Okamura} S.,  {Sasaki} T.,
  {Shioya} Y.,    {Taniguchi} Y.,  2006, \pasj, 58, 313

\bibitem[\protect\citeauthoryear{{Shioya}, {Taniguchi}, {Sasaki}, {Nagao},
  {Murayama}, {Saito}, {Ideue}, {Nakajima}, {Matsuoka}, {Trump}, {Scoville} \&
  {Sanders}}{{Shioya} et~al.}{2009}]{2009ApJ...696..546S}
{Shioya} Y.,  {Taniguchi} Y.,  {Sasaki} S.~S.,  {Nagao} T.,  {Murayama} T.,
  {Saito} T.,  {Ideue} Y.,  {Nakajima} A.,  {Matsuoka} K.,  {Trump} J.,
  {Scoville} N.~Z.,    {Sanders} D.~B.,  2009, \apj, 696, 546

\bibitem[\protect\citeauthoryear{{Spergel}, {Verde}, {Peiris}, {Komatsu},
  {Nolta}, {Bennett}, {Halpern}, {Hinshaw}, {Jarosik}, {Kogut}, {Limon},
  {Meyer}, {Page}, {Tucker}, {Weiland}, {Wollack} \& {Wright}}{{Spergel}
  et~al.}{2003}]{2003ApJS..148..175S}
{Spergel} D.~N.,  {Verde} L.,  {Peiris} H.~V.,  {Komatsu} E.,  {Nolta} M.~R.,
  {Bennett} C.~L.,  {Halpern} M.,  {Hinshaw} G.,  {Jarosik} N.,  {Kogut} A.,
  {Limon} M.,  {Meyer} S.~S.,  {Page} L.,  {Tucker} G.~S.,  {Weiland} J.~L.,
  {Wollack} E.,    {Wright} E.~L.,  2003, \apjs, 148, 175

\bibitem[\protect\citeauthoryear{{Springel}}{{Springel}}{2005}]{Springel05}
{Springel} V.,  2005, \mnras, 364, 1105

\bibitem[\protect\citeauthoryear{{Stark}, {Ellis}, {Richard}, {Kneib}, {Smith}
  \& {Santos}}{{Stark} et~al.}{2007}]{2007ApJ...663...10S}
{Stark} D.~P.,  {Ellis} R.~S.,  {Richard} J.,  {Kneib} J.,  {Smith} G.~P.,
  {Santos} M.~R.,  2007, \apj, 663, 10

\bibitem[\protect\citeauthoryear{{Stenflo}}{{Stenflo}}{1976}]{1976A&A....46...%
61S}
{Stenflo} J.~O.,  1976, \aap, 46, 61

\bibitem[\protect\citeauthoryear{{Trenti}, {Smith}, {Hallman}, {Skillman} \&
  {Shull}}{{Trenti} et~al.}{2010}]{2010ApJ...711.1198T}
{Trenti} M.,  {Smith} B.~D.,  {Hallman} E.~J.,  {Skillman} S.~W.,    {Shull}
  J.~M.,  2010, \apj, 711, 1198

\bibitem[\protect\citeauthoryear{{Verhamme}, {Schaerer} \&
  {Maselli}}{{Verhamme} et~al.}{2006}]{2006A&A...460..397V}
{Verhamme} A.,  {Schaerer} D.,    {Maselli} A.,  2006, \aap, 460, 397

\bibitem[\protect\citeauthoryear{{Wagner}, {M{\"u}ller} \&
  {Steinmetz}}{{Wagner} et~al.}{2008}]{2008A&A...487...63W}
{Wagner} C.,  {M{\"u}ller} V.,    {Steinmetz} M.,  2008, \aap, 487, 63

\bibitem[\protect\citeauthoryear{{Wang}, {Rhoads}, {Malhotra}, {Dawson},
  {Stern}, {Dey}, {Heckman}, {Norman} \& {Spinrad}}{{Wang}
  et~al.}{2004}]{2004ApJ...608L..21W}
{Wang} J.~X.,  {Rhoads} J.~E.,  {Malhotra} S.,  {Dawson} S.,  {Stern} D.,
  {Dey} A.,  {Heckman} T.~M.,  {Norman} C.~A.,    {Spinrad} H.,  2004, \apjl,
  608, L21

\bibitem[\protect\citeauthoryear{{Wannier}, {Lichten} \& {Morris}}{{Wannier}
  et~al.}{1983}]{1983ApJ...268..727W}
{Wannier} P.~G.,  {Lichten} S.~M.,    {Morris} M.,  1983, \apj, 268, 727

\bibitem[\protect\citeauthoryear{{Zheng}, {Cen}, {Trac} \&
  {Miralda-Escud{\'e}}}{{Zheng} et~al.}{2010}]{2010ApJ...716..574Z}
{Zheng} Z.,  {Cen} R.,  {Trac} H.,    {Miralda-Escud{\'e}} J.,  2010, \apj,
  716, 574

\end{thebibliography}

\appendix

\section{Lyman-$\alpha$ Radiative Transfer}
\label{sec:montecarlo}

The basic principle of \clara is to follow the individual scatterings of
a \lya photon as it travels through a distribution of gaseous hydrogen.
Each scattering, which is in fact an absorption and re-emission, does
not modify the frequency of the photon in the rest-frame of the hydrogen
atom. But due to the peculiar velocities of the atom in the new
direction of propagation of the photon, the new frequency in a
laboratory rest-frame is different from the incoming frequency. Thus the
photon performs a random walk not only in space but also in frequency.  

The relevant properties of the gaseous hydrogen can be fully described
by its density, temperature and bulk velocity. This is sufficient to
describe the emergent spectra of a source of \lya photons embedded in
gaseous hydrogen. It is important to observe that none of the emitted
photons is completely lost by absorption as it is immediately
re-emitted. The original spectrum morphology of the \lya source is
modified, but the only way to lose the energy input by a \lya source is
through dust absorption. A simple description of the dust abundance in
the gas must then be included to calculate its effect on the outgoing
properties of the traveling \lya photons. 

In the next subsections we give a detailed account on the basic
underlying physics of the qualitative description we have just given.
Once the physical fundamentals are described, we describe how these are
implemented in the code. The last subsection is devoted to  show the
results available analytical test cases we applied the code to.

\subsection{Physical Principles}
\label{subsec:physical}

\subsubsection{Scattering}

The scattering cross section of a \lya photon is a function of the
photon frequency. In the rest-frame of the atom it is equal to 

\begin{equation}
\sigma(\nu) = f_{12} \frac{\pi e^2}{m_ec}\frac{\Delta\nu_L/2\pi}{(\nu-\nu_0)^2 + (\Delta\nu_L/2)^2}, 
\end{equation}
where $f_{12}=0.4162$ is the \lya oscillator strength, $\nu_{0}=2.466\times
10^{15}$Hz is the line centre frequency, $\Delta\nu_{L}=4.03\times
10^{-8}\nu_0=9.936\times 10^7$ Hz is the natural line width, and the others
symbols conserve their usual meaning.

In the case of a Maxwellian distribution of atom velocities, after
convolving the individual cross section with the atom velocity
distribution, we can write down the average cross section as 

\begin{equation}
\sigma(x)=\sigma_x = f_{12} \frac{\sqrt{\pi} e^2}{m_ec\Delta\nu_{D}} H(a,x),
\label{eq:sigma_x}
\end{equation}
where $H(a,x)$ is the Voigt function,

\begin{equation}
  H(a,x) = \frac{a}{\pi}\int_{-\infty}^{\infty}\frac{e^{-y^2}}{(x-y)^2 +a^2}dy,
\end{equation}
$\Delta\nu_{D}=(v_{p}/c)\nu_0$ is the Doppler frequency width, $v_{p} =
(2kT/m_{H})^{1/2}$ is $\sqrt{2}$ times the velocity dispersion of the Hydrogen
atom, $T$ is
the gas temperature, $m_{H}$ is the Hydrogen atom mass,
$a=\Delta\nu_L/(2\Delta\nu_D)$ is the relative line width and $x=(\nu_i -
\nu_0)/\Delta\nu_D$ is a re-parameterisation of the photon frequency respect to
the line centre normalised by the temperature dependent Doppler frequency width
of the gas. 

The scattering is coherent in the rest-frame of the photon, but to an
external observer, any motion of the atom will add a Doppler shift to
the photon. Measuring the velocity of the atom, ${\bf v}_a$, in units of
thermal velocity ${\bf u} = {\bf v}_a/v_{p}$, the frequency in the
reference frame of the atom is  

\begin{equation}
x^{\prime} = x- {\bf u}\cdot{\bf n}_i,
\label{eq:new_x}
\end{equation}
where ${\bf n}_i$ is a unit vector in the incoming direction of the photon. 

In the general case, the scattering of the \lya atom is not isotropic.
For symmetry reasons the scattering is isotropic in the azimuthal
direction, respect to the outgoing scattering direction. The
distribution of the scatter directions depends only on the angle,
$\theta$, between the incoming and outgoing direction of the photons,
${\bf n}_i$ and ${\bf n}_{o}$, respectively. 

The information on the outgoing angle $\theta$ is encoded in the phase
function, $W(\theta)$. In general the angular momenta of the initial,
intermediate and final states are involved in the calculation of
$W(\theta)$. In the case of resonant scatter the initial and final state
are the same. The intermediate state corresponds to the excited state.
Following the notation $n l_{J}$There are two possible excited states,
$2P_{1/2}$ and $2P_{3/2}$.  

Specifically, in the dipole approximation the phase function can be
written as: 

\begin{equation}
  W(\theta)\propto  1+\frac{R}{Q}\cos^2\theta
\label{eq:phase}
\end{equation}
where  $R/Q=0$ for the $2P_{1/2}\to 1S_{1/2}$ transition and $R/Q=3/7$ for the $2P_{3/2}\to 1S_{1/2}$ case.

The spin multiplicity of each sate is $2J+1$, meaning that the
probability of being excited to the $2P_{2/3}$ state is twice that of
the $2P_{1/2}$ state. These results are valid only in the core of the
line profile. \cite{1976A&A....46...61S} found that in the wings quantum
mechanical interference between the two lines act in such a way as to
give a scattering resembling a classical oscillator. In that case the
phase function takes the form  

\begin{equation}
  W(\theta)\propto  1+\cos^2\theta
\end{equation}

The travelling distance, $l$, of a \lya photon  of frequency $x$ can be expressed as

\begin{equation}
\tau_x = \sigma_s n_{H} l,
\label{eq:tau_H}
\end{equation}
where $n_{H}$ is the neutral hydrogen number density, and it has been assumed
that along the path $l$ the temperature and bulk velocity field of the gas are
constant, to ensure that the photon frequency can be represented by the same
value of $x$ along its trajectory.

In what follows, we will always characterise an homogeneous and static
medium using the optical depth $\tau_{0}$ at the line center. 

\subsubsection{Dust Absorption}
In the case of dust interaction the photon can either be scattered or
absorbed. The optical depth of dust, $\sigma_d$,  can be generally expressed
as the sum of an absorption cross section $\sigma_a$, and a scattering
cross section $\sigma_s$.  

\begin{equation}
  \sigma_d = \sigma_a + \sigma_s.
\end{equation}

The determination of these two cross sections can be achieved using two
different approaches. We name the first approach the \emph{ab initio}
approach. The ab initio approach seeks to determine the values of the
dust cross sections from individual studies of dust grain properties and
its interaction with photons. This is the approach used in the studies
by \cite{2006A&A...460..397V}. The second approach is a phenomenological
one and it defines the dust properties in relation to the gas properties
in the galaxy. The dust cross section properties are then derived from
observations. In the interest of keeping our model simple to operate and
with a good match to the level of detail required for the approximation,
we proceed with an ab initio approach.

We express then the absorption and scattering cross sections as

\begin{equation}
\sigma_{a,s}  = \pi d^2 Q_{a,s},
\end{equation}
where $d$ is represents an average dust radius and $Q_{a,s}$ is an
absorption/scattering efficiency. The dust albedo can be then expressed as  

\begin{equation}
A = \frac{Q_{s}}{Q_{a} + Q_{s}}.
\label{eq:albedo}
\end{equation}

At UV wavelengths the emission and absorption processes are equally
likely, with $Q_{a}\sim Q_{s}\sim 1$, making the dust albedo around
$\sim 1/2$. We now express the dust of optical depth, $\tau_{d}$ in an
analogous way to the neutral hydrogen optical depth Eq.(\ref{eq:tau_H})
for a parcel of dust of linear dimensions $l$

\begin{equation}
\label{eq:tau_dust}
\tau_{d} = \sigma_d n_{d} l,
\end{equation}
where $n_{d}$ represents the number density of dust particles and it has been
assumed again that the dust cross section and dust number density are constant
on the scale of $l$. 

\subsection{The Radiative Transfer Code}

The code implements a Monte Carlo approach to the radiative transfer.
\clara follows the successive scattering of individual photons as they
travel through the gas distribution, changing at each scatter the
direction of propagation and frequency of the photon. We describe now in
detail the technical implementation. 

\subsubsection{Initial Conditions}
The problem to solve defines the physical characteristics of the gas
distribution and the initial conditions of the \lya emitted photons. The gas is
described by the following characteristics:

\begin{itemize}
\item{The geometry of the gas distribution. In this section we will
    present results for the following configurations: infinite slab and sphere} 
\item{The size of the gas
    distribution. This is parameterised by the hydrogen optical depth,
    $\tau_{0}$ at line centre $x=0$.  In all the geometries we measure the optical depth from
    the centre of the configuration to its nearest border.} 
\item{The temperature of the gas distribution, $T$. This is set to a
    constant for all the gas distributions explored in this paper.} 
\item{The gas bulk velocity field, ${\bf v_b}({\bf r})$. This is in
    general dependent on the position. The only bulk velocity field explored
    in this work corresponds to a Hubble like flow in the spherical geometries.} 
\item{The dust optical depth, $\tau_{d}$.} 
\item{The dust albedo, $A$.} 
\end{itemize}

The photons are described by the following properties:

\begin{itemize}
\item{The spatial distribution with respect to the gas distribution. There are
    two possibilities. All the photons are emitted from the centre of the gas
    distribution or they are homogeneously distributed throughout the gas
    volume.} 
\item{The initial direction of propagation. We assume that the emission is
    isotropic (in the local comoving frame).} 
\item{The initial frequency, $x_i$. We consider that all photons are
    injected in the centre of the line $x_{i}=0$.} 
\end{itemize}

The number of photons used in order to reach convergence are $N_{ph}\sim
5\times 10^4$. In some cases convergence is reached with $N_{ph}\sim 1
\times 10^4$ photons.

\subsubsection{Photon Displacement and Interaction}

For each photon of frequency $x$ and direction of propagation ${\bf
n}_{i}$, we determine the distance freely travelled until the next
interaction.  This optical depth is determined by sampling the
probability distribution $P(\tau) = \exp(-\tau)$ by setting $\tau_{s} =
-\ln (\rand)$, where \rand is a represents a random number between
$0\leq\rand<1$ following  an uniform distribution.

This optical depth interaction fixes the travel distance $l_{s}$ by:

\begin{equation}
l_{s} = \frac{\tau_{s}}{\sigma_x n_{H} + \sigma_d n_{d}}
\end{equation}

The photon travels a distance $l_{s}$ in the direction ${\bf b}_{i}$, at this
point we decide if the photon interacts either with an hydrogen atom or with a
dust grain. The probability of being scattered by a Hydrogen atom is 

\begin{equation}
P_H = \frac{\sigma_x n_{H}}{\sigma_x n_{H} + \sigma_d n_{d}}.
\end{equation}

Another random number $\rand$  is generated and compared to $P_{H}$. If
$\rand<P_{H}$ the photon interacts with Hydrogen, otherwise it interacts
with the dust grain. In the case of interaction with dust a new random
number is generated and compared with the albedo $A$. If the random
number is less than the albedo, the photon is scattered coherently in a
random direction, otherwise the photon is absorbed and not considered
for further scatterings during the rest of the simulation.  

In the case of photon interaction, the situation is more involved. As
already described by Eq. (\ref{eq:new_x}), the new photon frequency in
the observer frame depends on the specific velocity of the atom, ${\bf
u}$. It is therefore necessary to draw a velocity for the hydrogen atom.
The preferred velocity for the atom cannot be drawn from a isotropic
Gaussian distribution, there is an implicit spatial anisotropy in this
case. The opacity of the gas is frequency dependent: the preferred atom
velocity in the parallel direction to the photon  propagation is
different from the velocity in the perpendicular direction.  

The atom velocity is thus determined in two steps. In the first step the
two perpendicular velocities are selected from a Gaussian distribution.
In the second step the parallel component of the atom velocity is
calculated. The parallel component is calculated from a distribution
calculated as the convolution of a Gaussian (representing the intrinsic
velocity) convolved with the Lorentzian probability of the atom being
able to scatter the photon: 

\begin{displaymath}
f(u_{||}) \propto {\mathcal G}(u_{||}) \times {\mathcal L}(x-u_{||}) \propto {e^{-u_{||}^2}}  \times \frac{a}{\pi}\frac{1}{(x-u_{||})^2 + a^2}.
\end{displaymath}
\noindent
The resulting normalised probability is:

\begin{equation}
f(u_{||}) = \frac{a}{\pi H(a,x)}\frac{e^{u_{||}^2}}{(x-u_{||})^2 + a^2}.
\label{eq:u_parallel}
\end{equation}

The distribution represented by  Eq. (\ref{eq:u_parallel}) is not
analytically integrable, therefore we generate the parallel velocities
$u_{||}$ by means of the rejection method.  

We stress that we do not use any of the commonly used speeding
mechanisms. These speeding techniques are motivated by the physical fact
that scattering of atoms in the core frequencies are irrelevant from the
point of view of the change in frequency and position that they
represent. Given the uncertainty of applying these techniques (despite
of careful code calibration) in the presence of dust, we decide to use
the simplest rejection method. 

The new emission direction of propagation (in the atom rest-frame),
${\bf n}_f$ is determined as already described in detail in subsection
\ref{subsec:physical}. In the observers frame, the final frequency is:   

\begin{equation}
x_{f} = x_{i} - {\bf n}_{i}\cdot {\bf u} + {\bf n}_{f}\cdot {\bf u} + g(1- {\bf n}_{i}\cdot {\bf n}_f),
\end{equation} 

\noindent
where the factor $g$ accounts for the recoil effect.

\begin{figure*}
\begin{center}
\includegraphics[width=0.30\textwidth]{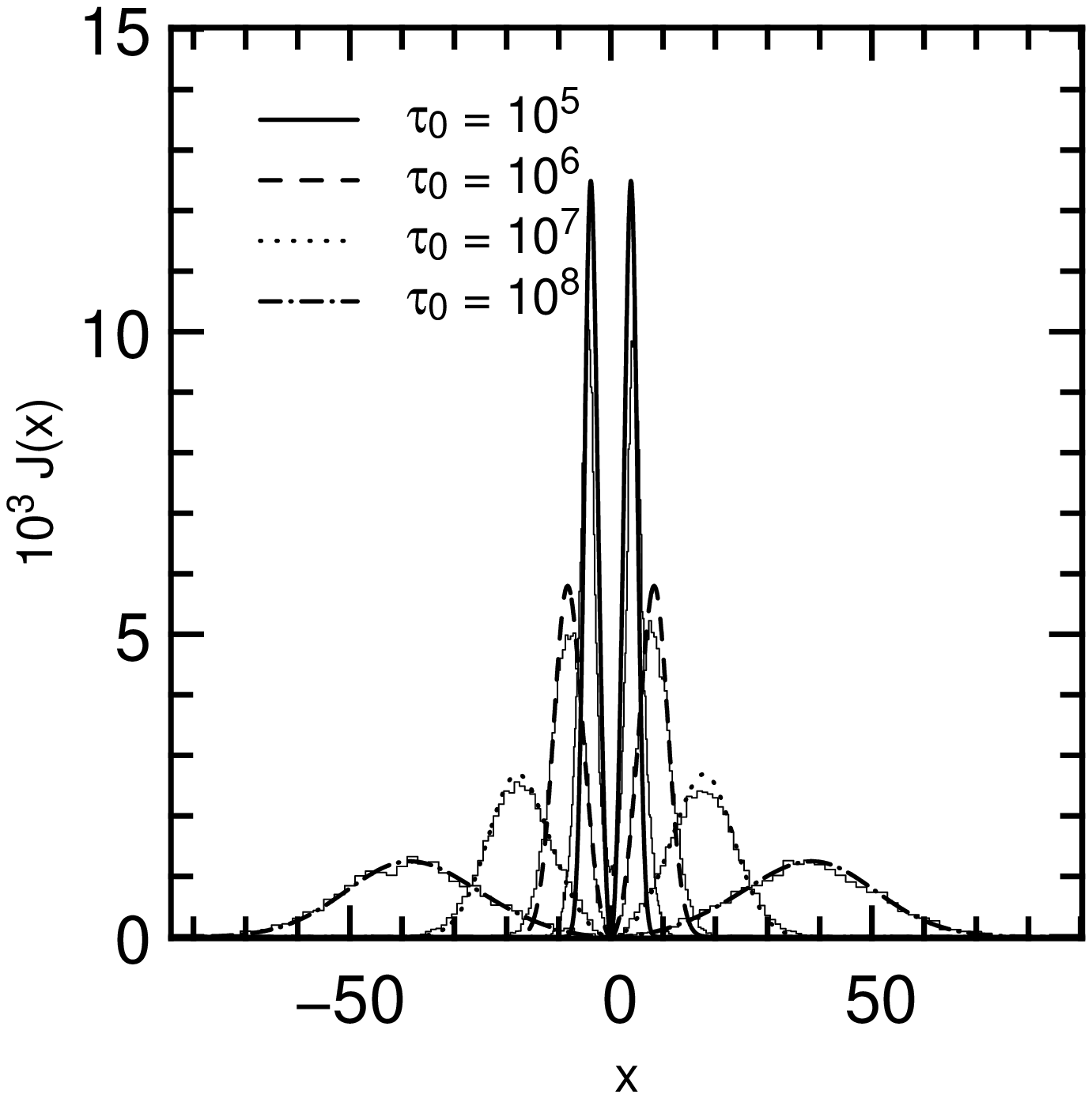}\hspace{0.6cm}
\includegraphics[width=0.30\textwidth]{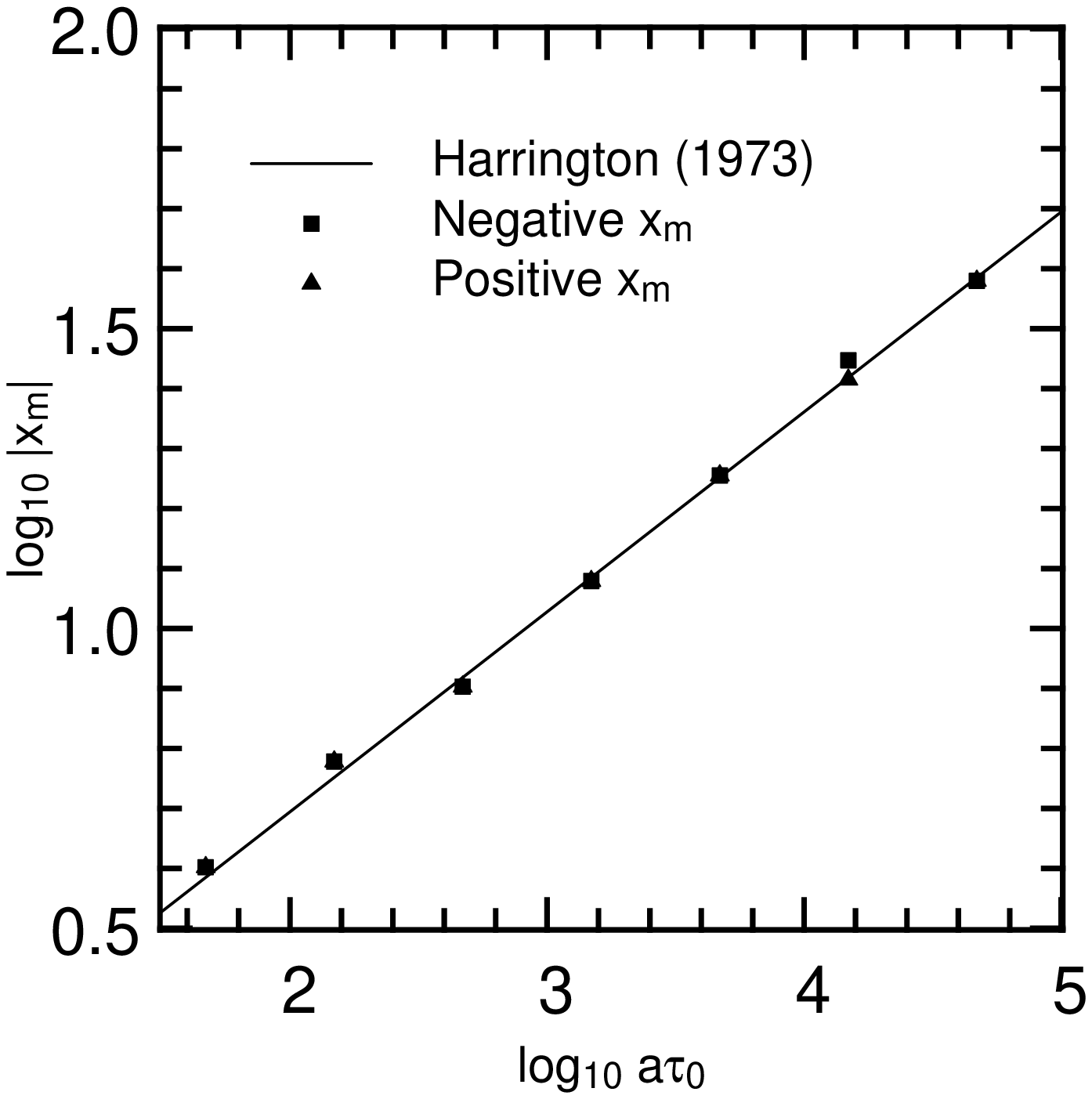}\hspace{0.6cm}
\includegraphics[width=0.28\textwidth]{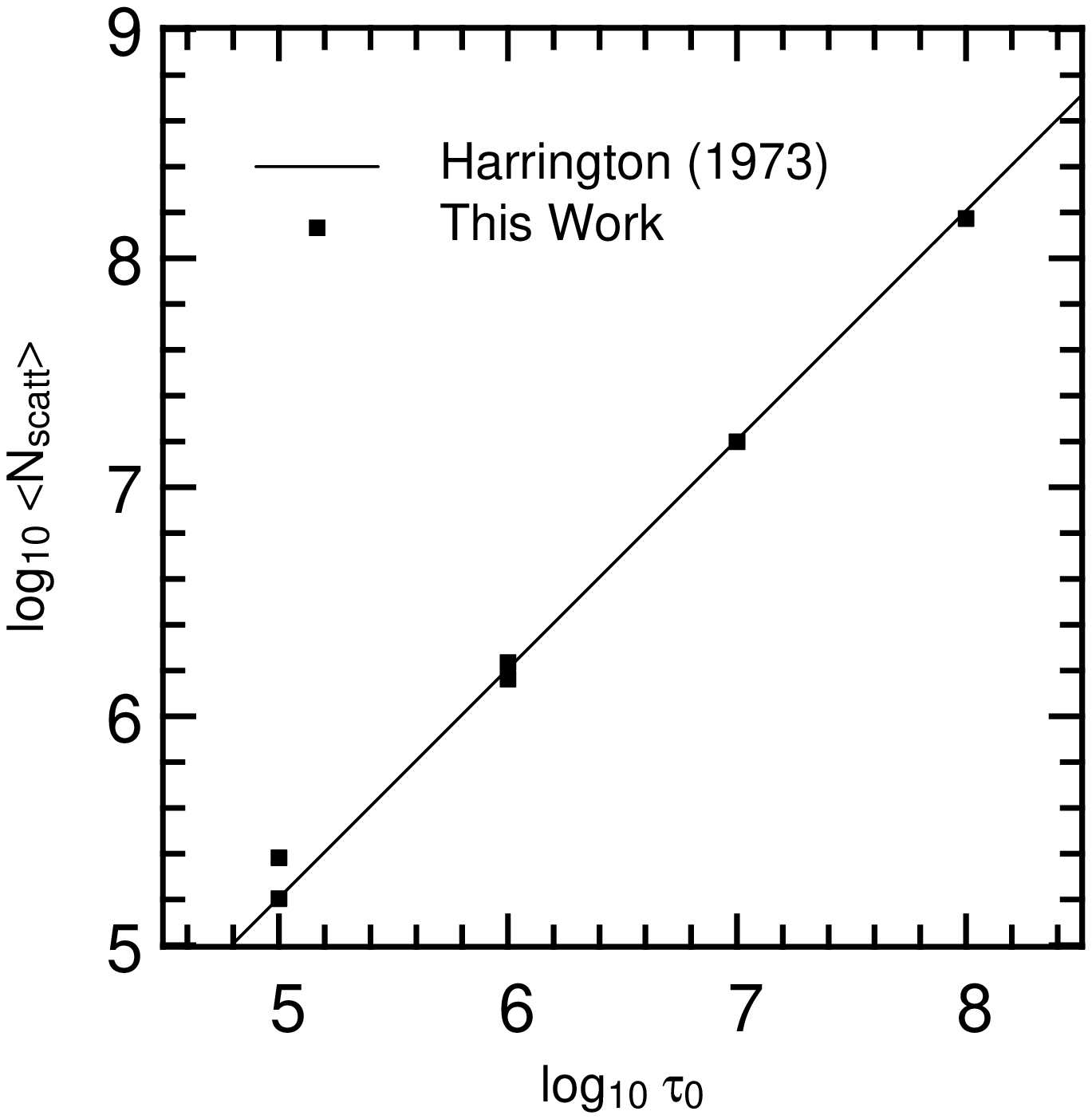}
\end{center}
\caption{Analytical tests related to the infinite homogeneous gas slab. The
  left panel show the outgoing spectra for different hydrogen optical depths,
  compared to the expected analytical solutions. The middle panel compares
  the positions of the maxima in the outgoing spectra to the expected
  analytical estimates for a series of 5 different runs. The panel on the
  right compares the average number of scatterings the \lya photons go through
  before escaping the slab. \clara passes successfully these three tests.}
\label{fig:neufeld}
\end{figure*}

\subsection{Analytical Tests}

\begin{table*}
\begin{center} 
\begin{tabular}{cccccccc}\hline\hline
$T (K) $ & $\tau_0$ & $V_{max}$& $\tau_{a}=(1-A)\tau_d$ 
& $f_{esc,slab}^{C}$ & $f_{esc,slab}^{H}$& $f_{esc,sph}^C$ &
$f_{esc,sph}^H$ \\\hline
$10^4$ & $10^6$ & 0 &$5.0\times 10^{-5}$ & 0.998 & 0.999& 0.999 & 0.999 \\
$10^4$ & $10^6$ & 0 &$5.0\times 10^{-4}$& 0.991& 0.993 &0.995 & 0.997 \\
$10^4$ & $10^6$ & 0 &$5.0\times 10{-3}$ & 0.917 & 0.937&0.957& 0.976 \\
$10^4$ & $10^6$ & 0 &$5.0\times 10^{-2}$ & 0.471 & 0.590&0.664& 0.803\\
$10^4$ & $10^7$ & 0 &$5.0\times 10^{-5}$ & 0.998 & 0.998&0.999& 0.999 \\
$10^4$ & $10^7$ & 0 &$5.0\times 10^{-4}$ & 0.982 & 0.987&0.991& 0.995 \\
$10^4$ & $10^7$ & 0 &$5.0\times 10^{-3}$ & 0.853 & 0.894&0.925& 0.962 \\
$10^4$ & $10^7$ & 0 &$5.0\times 10^{-2}$& 0.291 & 0.453&0.493& 0.720 \\
$10^4$ & $10^7$ & 0 &0.5 & 0.0029 & 0.084&0.014& 0.224 \\
$10^4$ & $10^7$ & 0 &0.1 & 0.127 & 0.295&0.280& 0.568 \\
$10^4$ & $10^7$ & 0 &0.2 & 0.036 & 0.179&0.111& 0.407\\
$10^4$ & $10^7$ & 0 &0.4 & 0.0069 & 0.103&0.027& 0.265\\
$10^4$ & $10^7$ & 0 & 1.0 & 0.00023 & 0.049 &0.0011& 0.136\\ 
$10^4$ & $10^7$ & 0 & 4.0 & $<$ 0.0001& 0.014 & $<$0.0001&0.042\\ 
$10^4$ & $10^7$ & 0 & 5.0 & $<$ 0.0001& 0.012& $<$0.0001&0.036\\ 
$10^4$ & $1.2\times 10^7$ & 0 & 0.5 & - & -&0.012 & -\\ 
$10^4$ & $1.2\times 10^7$ & 20 & 0.5 & - & -&0.017 & -\\ 
$10^4$ & $1.2\times 10^7$ & 200 & 0.5  & - & -&0.140 & -\\ 
$10^4$ & $1.2\times 10^7$ & 2000 & 0.5 & - &-&0.283 & -\\ \hline
\end{tabular}
\caption{Summary of the physical characteristics for all runs, both with
  infinite slab and spherical geometry of a dusty gas distribution. $V_{max}$ refers to the
  parameterisation of the radial velocity profile imposed on the sphere
  following a Hubble-like law. The escape fraction in the case  of centrally
  distributed sources in a sphere is denoted by $f_{esc,sph}^C$, while   $f_{esc,sph}^H$
  corresponds to the escape fraction when the \lya sources are   homogeneously
  distributed. The escape fractions for the infinite slab are denoted
  by $f_{esc,slab}^{C}$ and $f_{esc,slab}^{H}$.} 
\label{table:runs_slab}
\end{center}
\end{table*}

The basic tests on \clara are based on the available analytical
solutions for a thick, plane-parallel, isothermal infinite gas slab of
uniform density.  The solution of the infinite slab provides the
analytical expressions for: the distribution of outgoing frequencies,
the average number of scatterings and the position of the peaks of the
outgoing spectrum. In the case of a dusty infinite slab one we also have
an analytical expression for the escape fraction
\citep{1973MNRAS.162...43H,1991ApJ...370L..85N}. 

The expression for the analytic emergent spectrum as a function of
frequency shift for a mid-plane source reads:

\begin{equation}
J(x)   = \frac{\sqrt{6}}{24}\frac{1}{a\tau_{0}} \frac{x^2}{\cosh(\sqrt{\pi^4/54} |x^3-x_{i}^3|/a\tau_{0})},
\label{eq:neufeld}
\end{equation}
where $\tau_{0}$ is the optical depth measured from the mid-plane to the slice
boundary and $x_{i}$ is the frequency of the photons injected in the
slice. For $x_{i}=0$ the emergent spectrum has a symmetric two-peaked
shape. 

Setting $\partial J/\partial x=0$ \citep{1973MNRAS.162...43H}, it can be
shown that the position of the maxima is: 

\begin{equation}
x_m = \pm 1.066 (a\tau_{0})^{1/3}, 
\label{eq:maxima}
\end{equation}
while the average number of scatterings is:

\begin{equation}
\avg{N_{s}} = 1.612\tau_{0}.
\label{eq:n_scatter}
\end{equation}

If dust is added to the gas distribution, it is then possible from
analytical considerations to estimate the scalings of the escape
fraction from the slab physical properties. The expression for the
escape fraction reads

\begin{equation}
f_{esc} = \frac{1}{\cosh \left( \xi^{\prime} \sqrt{(a\tau_{0})^{1/3}\tau_a}\right)},
\label{eq:f_esc}
\end{equation}
where $\xi^{\prime}=\xi\sqrt{3}/\pi^{5/12}$ and $\xi=0.565$ is a parameter found from a fit.

In the following we compare the analytical results described by
Eqs.(\ref{eq:neufeld}) to (\ref{eq:f_esc}) with the results from \clara.
In Figure \ref{fig:neufeld} we show the outgoing spectra for the
infinite slab with four different optical depths $\tau_0 = 10^{5}$,
$10^6$, $10^7$ and $10^8$, and a constant temperature of $T=10$K. The
broad shape of the double peak is reproduced. The agreement with the
expected analytical solution becomes better as the hydrogen optical
depth increases, as expected. The simulation and the analytical result
are expected to hold for optical depths and temperatures such as
$a\tau_0 > 10^3$. 

In the middle panel of Figure \ref{fig:neufeld}, we have the positions
of the positive and negative peaks in the outgoing spectra, as a
function of the product $a\tau_0$. These results were obtained from 7
different runs with their characteristics listed in Table
\ref{table:runs_slab}. In the right panel we compare the number of
scatterings with the theoretical prediction, which is only dependent on
the hydrogen optical depth $\tau_{0}$. In both cases we find optimal
agreement with the theoretical expectations. We stress again that even
though we reproduce the right number of scatterings, we have not
implemented any acceleration scheme based on the skip scattering scheme.

\begin{figure}
\begin{center}
\includegraphics[width=0.40\textwidth]{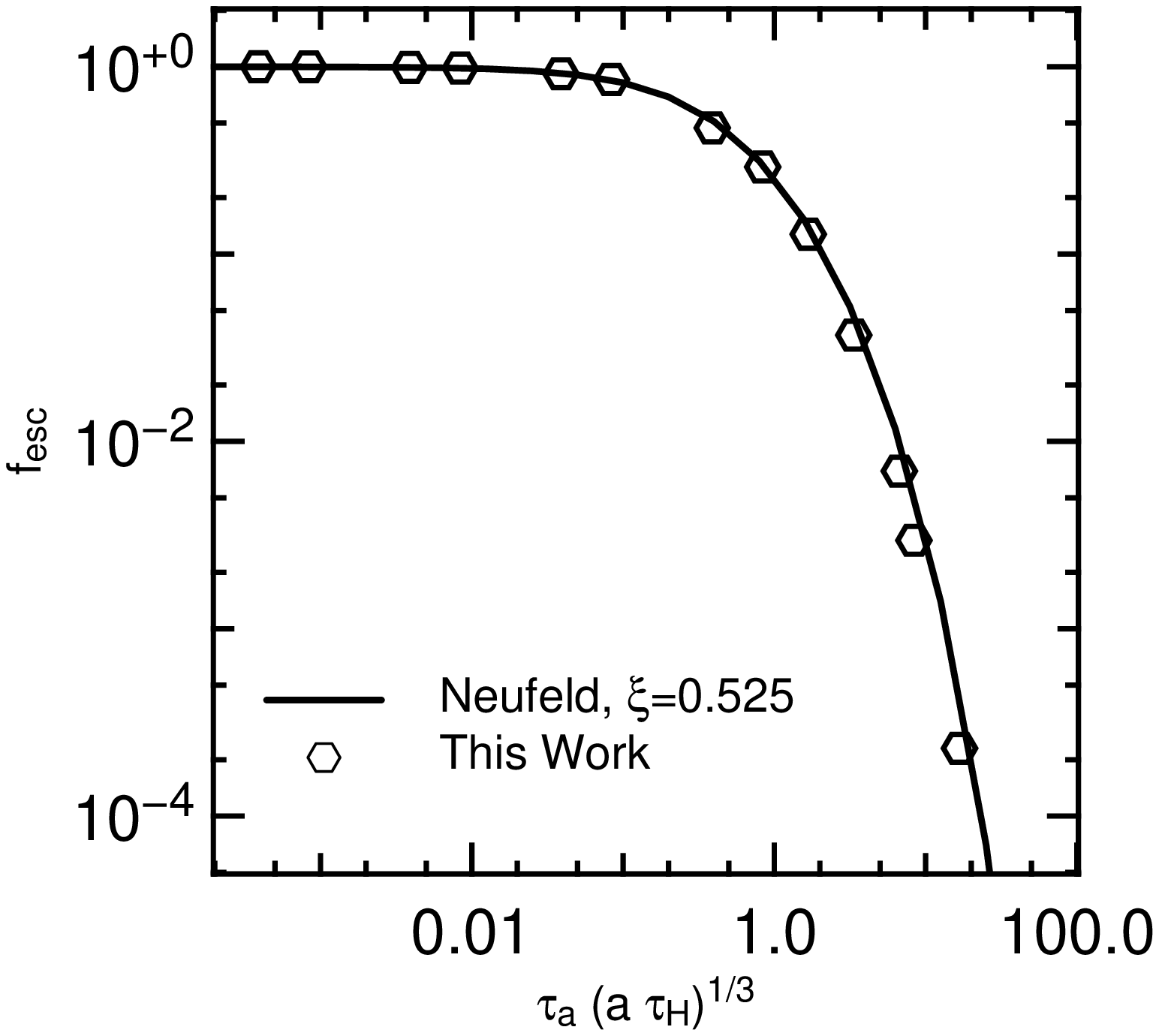}
\end{center}
\caption{Escape fraction for \lya\ photons emitted from the centre of a
  infinite slab of gas. The hexagons represent the results of {\texttt{CLARA}}, the continuous line, the analytical solution
  of Eq.(\ref{eq:f_esc}).}
\label{fig:escape_slab}
\end{figure}

In Figure \ref{fig:escape_slab} we show the results concerning the
escape fraction for the dusty infinite slab. We compare our results to
the expected result from Eq. (\ref{eq:f_esc}). In Table
\ref{table:runs_slab} we list the parameters for the dusty slab run, as
well as the values of the escape fraction we find. The theoretical
expectations for the escape fraction take the interaction with dust as
an absorption. In the case of our runs, as we consider an albedo of
$A=0.5$, the optical depth of dust grains has to be replaced by an
effective value of an absorbing material $\tau_{a} = (1-A)\tau_d  =
0.5\tau_d$, where $\tau_d$ is defined by Eq. (\ref{eq:tau_dust}). Once
again, we find a reasonable agreement with the theoretical expectations.  

At this point we are confident that \clara provides the expected
outgoing spectra and number of scatterings. The dust implementation has
been tested against the available theoretical constraints, yielding
satisfactory results.

The case of spherical symmetry offers analytical solutions in specific
cases \citep{2006ApJ...649...14D}. Besides, the case of a expanding
(contracting) sphere has been extensively studied in the literature with
different Monte Carlo codes. The spherical symmetry situations gives us
the possibility to perform further tests on {\texttt{CLARA}}. We will
deal with the case of a static, isothermal and homogeneous sphere of gas
first. Subsequently we will add a Hubble-like flow to the sphere.

\begin{figure}
\begin{center}
  \includegraphics[width=0.40\textwidth]{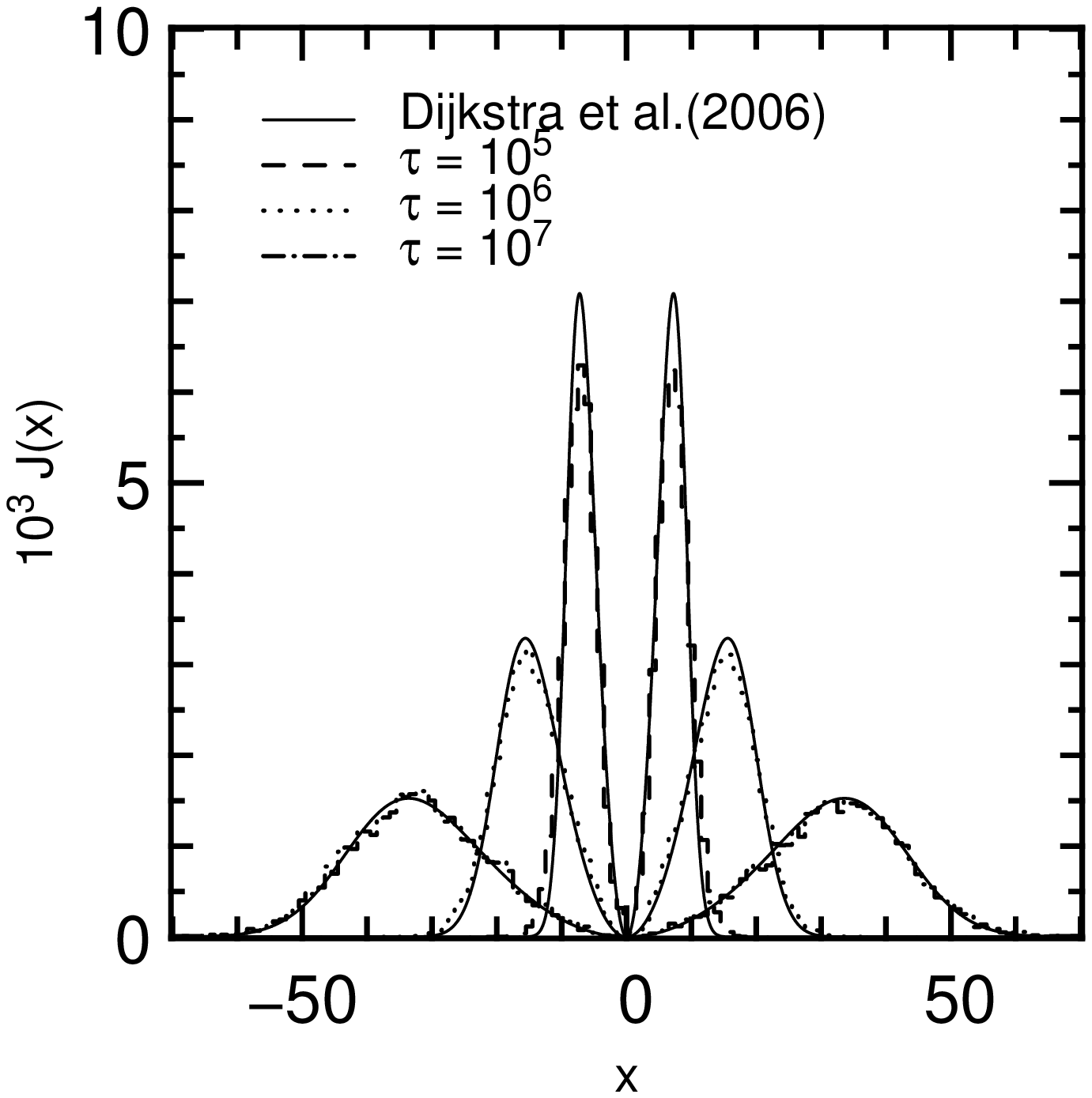}
\end{center}
\caption{Emergent line profile for the static and dustless sphere test. The
  photons are emitted at the centre of a sphere with different hydrogen
  optical depths. The analytical solution follows \citet{2006ApJ...649...14D}.}
\label{fig:sphere}
\end{figure}

\begin{figure}
\begin{center}
\includegraphics[width=0.40\textwidth]{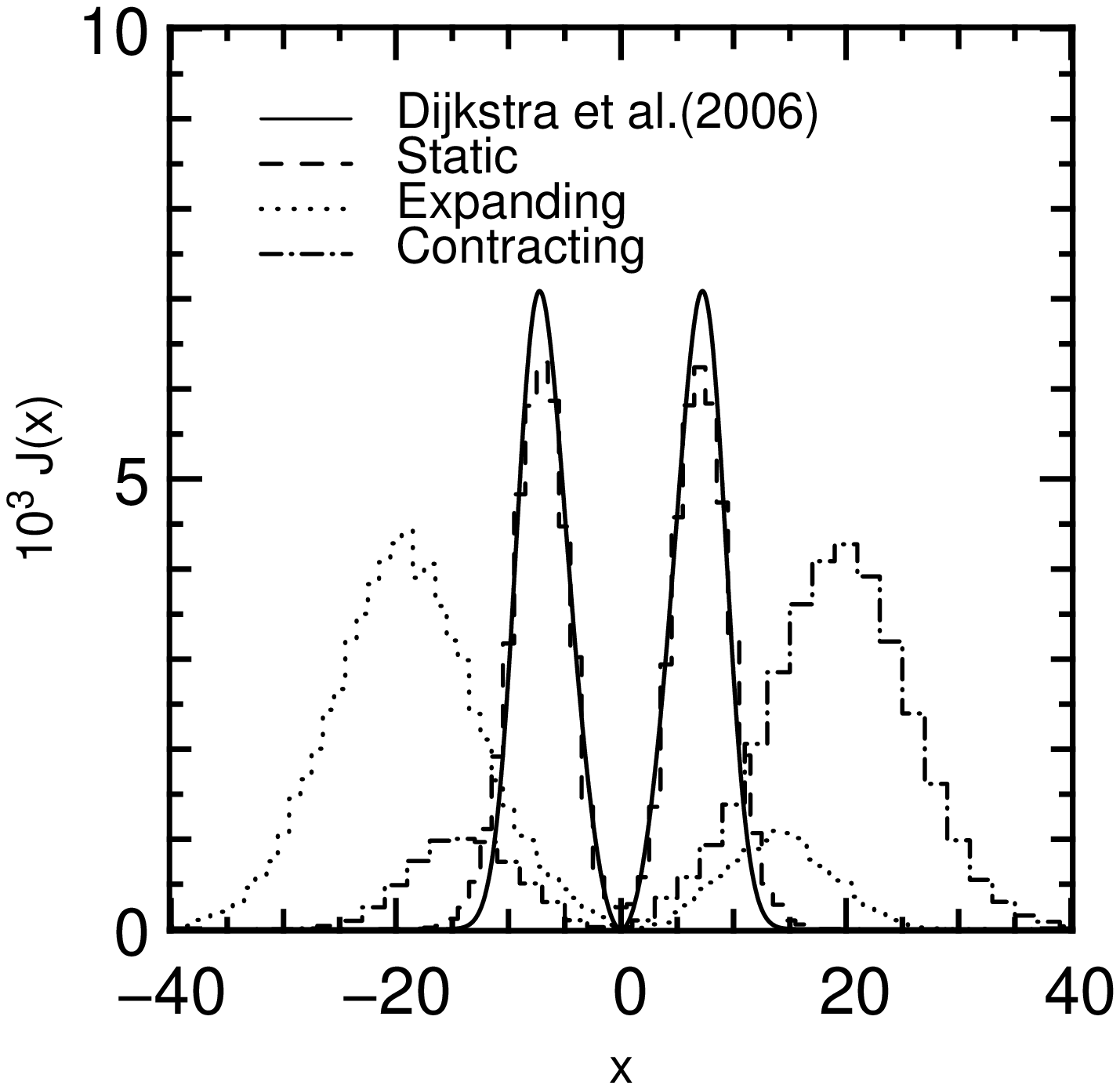}
\end{center}
\caption{Emergent line profile in the collapsing/outflowing sphere tests. The
  collapsing/outflowing solutions are symmetric under parity transformations 
  $x\leftrightarrow -x$.}
\label{fig:sphere_hubble}
\end{figure}

\cite{2006ApJ...649...14D} computed the analytic solution for the
emergent spectrum of a point-like   \lya source surrounded by an
homogeneous, static gas distribution at a constant temperature. The
solution can be expressed with the same functional form as the
Harrington-Neufeld solution
\citep{1973MNRAS.162...43H,1991ApJ...370L..85N}.

We setup different configurations in \clara for a series of optical
depth and temperatures. Figure \ref{fig:sphere} shows \clara outputs
compared with the corresponding analytical solutions. Once more, in the
limit of large values for $a\tau_{0} > 10^3$ we recover the behaviour
expected by the analytical solution provided by
\cite{2006ApJ...649...14D}.

We add to the homogeneous sphere a radial Hubble like flow as a function
to the distance to the centre of the sphere 

\begin{equation}
V_r(r) = V_{max}\frac{r}{R}, 
\end{equation}
where $V_{max}$ is a constant which can be selected to be positive or negative. 

The outgoing spectra of this type of configuration has not been worked
out from an analytical point of view. However, it is part of the most
studied physical setups by Monte Carlo codes similar to ours. The
quantitative behaviour of our results follows the behaviour of
previously published results. In the expanding case, there is
suppression of the blue peak accompanied by an enhancement in the red
wing of the emergent spectrum. The case of the collapsing sphere
displays the enhancement in the blue peak and suppression in the red
peak.

\subsection{The Homogeneous Source Distribution}

\begin{figure}
\begin{center}
\includegraphics[width=0.40\textwidth]{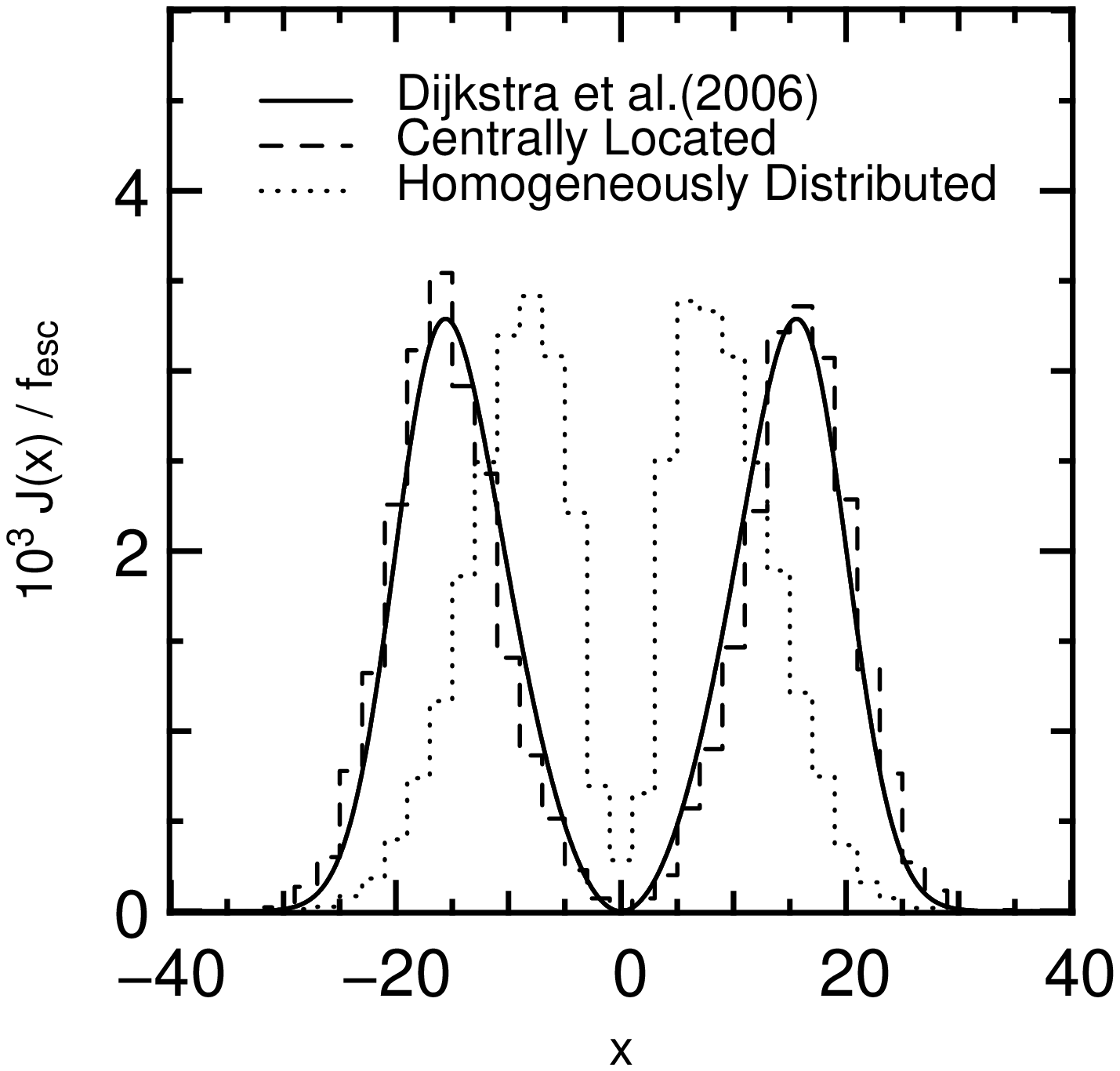}
\end{center}
\caption{Emergent line profile in the case of a static dusty sphere. The
  parameters for the run where $T=10^4$K, $\tau_{0}=10^7$ and $\tau_d=0.4$. The
  profiles are corrected for the escape fraction. The different line profiles
  correspond to the case of centrally distributed sources versus homogeneously
  distributed. The case of centrally distributed sources is compatible with
  the analytic solution of \citet{2006ApJ...649...14D}. The case of
  homogeneously distributed sources presents a different line profile. The
  maxima are closer to the line center, and each peak is asymmetric respect to
  the maximum. }
\label{fig:sphere_homo}
\end{figure}

The model of the LAEs suggested in the paper is based on the
approximation of an homogeneous dusty gas sphere, with the photon
sources distributed homogeneously inside the sphere. In the body of the
paper we explored in detail the consequences of this model on the escape
fraction. At the same time we have studied the effect on the outgoing
spectra.

Figure \ref{fig:sphere_homo} presents the outgoing spectra already
corrected by the escape fraction.  The dashed line shows the result for
the sphere with sources located at the center, the solid line is the
expected result for the case of the dustless sphere. The match between
the two curves shows that effectively the shape of the outgoing spectra
is preserved in the presence of dust. This result is expected because
all the photons are emitted at the same location and experience the same
optical depth.

In case of homogeneously distributed sources, the spectral shape is
different. The peaks are located closer to the line centre due to the
fact that now, on average, the photons experience a smaller optical
depth. At the same time, each wing presents now an asymmetrical shape
with respect to these maxima. Furthermore, there is now some amount of
photons very close to the center of the line, \emph{i.e.} emitted close
to the surface, that escaped without any scatter.

We have presented {\texttt{CLARA}}, a radiative transfer code for \lya
radiation. The code propagates \lya photons through arbitrary
distributions of hydrogen density, temperature, velocity structures and
dust distributions. The code has successfully passed all the available
tests for codes of its kind, and has allowed us to study simplified
geometries in the present paper.

\end{document}